\documentclass[iop]{emulateapj}

\usepackage{color}

\slugcomment{Accepted for publication in ApJ}

\shorttitle{Broad-line AGN BHMF and ERDF at z=1.4}
\shortauthors{Nobuta et al.}

\begin{document}

\title{Black Hole Mass and Eddington Ratio Distribution Functions 
of X-ray Selected Broad-line AGNs at $z\sim1.4$ in the Subaru XMM-Newton Deep Field\footnotemark[$*$]}

\author{
K.Nobuta\altaffilmark{1}, M.Akiyama\altaffilmark{1}, Y.Ueda\altaffilmark{2},
M.G.Watson\altaffilmark{3}, J.Silverman\altaffilmark{4}, K.Hiroi\altaffilmark{2},
K.Ohta\altaffilmark{2}, F.Iwamuro\altaffilmark{2}, K.Yabe\altaffilmark{2,5},
N.Tamura\altaffilmark{4,5},
Y.Moritani\altaffilmark{2,6}, M.Sumiyoshi\altaffilmark{2}, N.Takato\altaffilmark{5},
M.Kimura\altaffilmark{4,5}, T.Maihara\altaffilmark{2}, 
G.Dalton\altaffilmark{7,8}, I.Lewis\altaffilmark{7}, D.Bonfield\altaffilmark{7},
H.Lee\altaffilmark{7,9}, E.Curtis Lake\altaffilmark{7,10}, E.Macaulay\altaffilmark{7}, 
F.Clarke\altaffilmark{7}, K.Sekiguchi\altaffilmark{5}, C.Simpson\altaffilmark{11},
S.Croom\altaffilmark{12}, M.Ouchi\altaffilmark{13,4}, H.Hanami\altaffilmark{14}, T.Yamada\altaffilmark{1}
}
\email{akiyama@astr.tohoku.ac.jp}

\footnotetext[$*$]{Based in part on data collected at Subaru Telescope, which is operated by the National Astronomical Observatory of Japan.}

\altaffiltext{1}{Astronomical Institute, Tohoku University, 6-3, Aramaki, Aoba-ku, Sendai, 980-8578, Japan}
\altaffiltext{2}{Department of Astronomy, Kyoto University, Kitashirakawa-Oiwake-cho, Sakyo-ku, Kyoto, 606-8502, Japan}
\altaffiltext{3}{X-ray Astronomy Group, Department of Physics and Astronomy, Leicester University, Leicester LE1 7RH, UK}
\altaffiltext{4}{Kavli Institute for the Physics and Mathematics of the Universe, The University of Tokyo, Kashiwa, 277-8583, Japan}
\altaffiltext{5}{National Astronomical Observatory of Japan, Mitaka, Tokyo 181-8588, Japan}
\altaffiltext{6}{Hiroshima Astrophysical Science Center, Hiroshima University, Higashi-Hiroshima, Hiroshima, 739-8526, Japan}
\altaffiltext{7}{Department of Astrophysics, University of Oxford, Kebe Road, Oxford, OX1 3RH, UK}
\altaffiltext{8}{STFC Rutherford Appleton Laboratory, Chilton, Didcot, Oxfordshire OX11 0QX, UK}
\altaffiltext{9}{McDonald Observatory, University of Texas at Austin, 1 University Station C1402, Austin, TX 78712, USA}
\altaffiltext{10}{Institute for Astronomy, University of Edinburgh, Royal Observatory, Edinburgh EH9 3HJ, UK}
\altaffiltext{11}{Astrophysics Research Institute, Liverpool John Moores University, Twelve Quays House, Egerton Wharf, Birkenhead CH41 1LD, UK}
\altaffiltext{12}{Sydney Institute for Astronomy, School of Physics, University of Sydney, NSW 2006, Australia}
\altaffiltext{13}{Institute for Cosmic Ray Research, The University of Tokyo, Kashiwa 277-8582, Japan}
\altaffiltext{14}{Physics Section, Faculty of Humanities and Social Sciences, Iwate University, 020-8550, Morioka, Japan}

\begin{abstract}
In order to investigate the growth of super-massive black holes (SMBHs), 
we construct the black hole mass function (BHMF) and Eddington ratio 
distribution function (ERDF) of X-ray-selected broad-line AGNs at $z\sim1.4$ 
in the Subaru {\it XMM-Newton} Deep Survey (SXDS) field. In this redshift 
range, a significant part of the accretion growth of SMBHs is thought to 
be taking place. Black hole masses of X-ray-selected broad-line AGNs are 
estimated using the width of the broad \ion{Mg}{2} line and the
3000{\AA} monochromatic luminosity. We supplement the \ion{Mg}{2} FWHM values
with the H$\alpha$ FWHM obtained from our NIR spectroscopic survey.
Using the black hole masses of broad-line AGNs at redshifts between 1.18 and 1.68, 
the {\it binned} broad-line AGN BHMF and ERDF are calculated 
using the $V_{\rm max}$ method. 
To properly account for selection effects that impact the binned estimates, 
we derive the {\it corrected} broad-line AGN
BHMF and ERDF by applying the Maximum Likelihood method, 
assuming that the ERDF is constant regardless of the black hole mass.
We do not correct for the non-negligible uncertainties in virial
BH mass estimates.
If we compare the {\it corrected} broad-line AGN
BHMF with that in the local Universe, the 
{\it corrected} BHMF at $z=1.4$ has a higher number density above $10^{8} M_{\sun}$ but a
lower number density below that mass range. The evolution may be
indicative of a down-sizing trend of accretion activity among the SMBH population. 
The evolution of broad-line AGN
ERDF from $z=1.4$ to 0 indicates that the fraction of broad-line AGNs
with accretion rate close to the Eddington-limit is higher at higher redshifts. 
\end{abstract}


\section{Introduction}

Since the discovery that super-massive black holes (SMBHs) sit at the centers of 
most massive galaxy in the local Universe \citep[e.g.,][]{kormendy95}, 
determining their formation history remains one of big challenges in astrophysics. 
It has been further determined that the mass of SMBHs correlates tightly with the physical 
properties of their host spheroids \citep[$M_{\rm BH}$ vs. $M_{\rm bulge}$ 
relations; e.g.,][]{magorrian98, marconi03, gultekin09}. Such a correlation
implies a physical connection between the growth histories of SMBHs and the
spheroidal components of galaxies \citep[e.g.][]{boyle98}.

Bolometric luminosities of AGNs reflect the mass accretion rates of their SMBHs,
therefore the luminosity function of AGNs and its cosmological evolution
reflects the growth history of SMBHs through accretion \citep{soltan82}. Cosmological evolution 
of AGN luminosity functions have been evaluated using various AGN 
samples \citep[e.g.][]{ueda03, silverman08, croom09, aird10, assef11, simpson12}. 
The number density evolution of AGNs in 
different luminosity bins shows that higher luminosity AGNs, i.e. QSOs,
have a peak at higher redshifts, the so-called "down-sizing" trend of
cosmological evolution of AGNs. The total amount of  accreted matter
estimated by integrating the luminosity functions over luminosity and redshift
roughly matches to the estimated mass density of SMBHs in
the local Universe \citep{yu02, marconi04, shankar09}, thus accretion is thought to be
the dominant mode of SMBH growth. Applying the continuity equation for 
the SMBH population, \citet{marconi04} evaluate the 
average growth curves of massive and less-massive SMBHs as a 
function of redshift. These results imply that SMBHs grow rapidly at
redshifts between 1 and 2, and more massive SMBHs grow more than less massive SMBHs in 
the earlier Universe , as expected from the 
"down-sizing" trend of the AGN luminosity function.

The luminosity of an AGN does not simply reflect the mass of its SMBH.
In the calculations of the black hole growth history, the Eddington ratio, 
i.e. the ratio between the observed accretion rate and the Eddington-limited 
accretion rate ($\lambda_{\rm Edd}$), is assumed to be constant in AGNs 
with different luminosities and redshifts. Assuming a constant Eddington 
ratio for AGNs, their luminosity directly corresponds to the mass of 
their SMBHs, and the mass-dependent growth history of SMBHs can be calculated.
However, recent evaluations of the Eddington-ratio distribution
of AGNs in the local universe show that AGNs have a wide range of
Eddington-ratio with no preferred value \citep{kauffmann09, schulze10}. Therefore, 
in order to 
quantitatively understand the accretion growth history of SMBHs, 
it is necessary to evaluate SMBH masses in any AGN sample for which the
cosmological evolution of the luminosity function is evaluated.

Tools are now available to measure black hole masses of broad-line AGNs.
Reverberation mapping of local broad-line AGNs
\citep[e.g.,][]{peterson04} 
reveals the scaling relationship between the 5100{\AA} monochromatic 
luminosity ($L_{\rm 5100}$) of the broad-line AGN and the size of its broad H$\beta$ emitting 
region \citep{kaspi00, kaspi05}. Utilizing this relationship, black 
hole masses of a large sample of broad-line AGNs can be 
estimated from their luminosities and broad H$\beta$ line 
widths ($\Delta v_{{\rm H} \beta}$) \citep[e.g.,][]{vestergaard06} from the relationship
$M_{\rm BH} = f \Delta v_{{\rm H} \beta} ^2 L_{\rm 5100} ^{0.5}$ under the assumption that the 
broad-line region is virialized \citep{peterson99}. The factor $f$ depends on the
dynamical structure of the broad-line region, and it 
is empirically determined by assuming that the black hole mass
obtained from the reverberation mapping method and the velocity dispersion of
its host bulge follow the $M_{\rm BH}$ vs. $\sigma_{\rm bulge}$ relation 
of local non-active galaxies \citep{onken04}.
This method has been extended to
black hole mass estimations using other broad lines, such
as \ion{Mg}{2} $\lambda\lambda 2796,2803${\AA} 
\citep{mclure02, vestergaard09}, \ion{C}{4} $\lambda\lambda 1548,1551${\AA} \citep{vestergaard02, vestergaard06}, 
and H$\alpha$ \citep{greene05},
and using luminosities at other wavelengths such as the 3000{\AA} 
monochromatic luminosity \citep{mclure02}, the H$\alpha$ line 
luminosity \citep{greene05}, and the hard X-ray luminosity \citep{greene10a}. 
These relationships are calibrated against the black hole mass 
estimated by the reverberation mapping or that from the single-epoch 
broad-line width of H$\beta$ and the monochromatic
luminosity. Using these extended methods,
black hole masses of broad-line AGNs at various redshifts can
be estimated from their single-epoch optical spectra. 

Applying black hole mass estimates from single-epoch spectra
to statistical samples of broad-line AGNs, black hole mass functions (BHMFs) 
of broad-line AGNs can be evaluated 
\citep{wang06, greene07, greene09, vestergaard08,
vestergaard09, schulze10, shen12, kelly12}.
In the local Universe, \citet{schulze10} derived the BHMF and
Eddington-ratio distribution function (ERDF) of 
broad-line AGNs detected in the Hamburg/ESO AGN survey. 
They corrected the effects of the flux limits of their survey in their evaluation of the
broad-line AGN
BHMF and ERDF, i.e. the fact that the low-mass end of the sample only covers high 
Eddington ratio AGNs, by assuming that the ERDF is constant 
regardless of the black hole mass and applying Maximum likelihood method. 
Hereafter we label BHMF and ERDF derived by
$V_{\rm max}$ method with {\it binned} and those corrected for the detection
limit by Maximum likelihood method with {\it corrected}.
The {\it corrected} broad-line AGN BHMF covering $M_{\rm BH}$ values between
$10^{6.0} M_{\sun}$ and $10^{9.5} M_{\sun}$ and $\lambda_{\rm Edd}$ down to $0.01$
shows rather steep decrease in number density as a function of mass with no
significant break in the mass range covered.
Their {\it corrected} broad-line AGN
ERDF shows a steep decline at the Eddington limit
and a steep increase in the number density down to 
$\lambda_{\rm Edd}$ of $0.01$, following power-law with index of $\sim -1.9$.

Cosmological evolution of the BHMFs of broad-line AGNs has also been examined using
large samples of broad-line AGNs from the Sloan Digital Sky Survey (SDSS)
using the  $V_{\rm max}$ method \citep{vestergaard08, vestergaard09}
and a Bayesian approach \citep{kelly10, shen12, kelly12}. 
\citet{kelly10}, \citet{shen12}, and \citet{kelly12} derive the cosmological evolution of the 
BHMF of broad-line AGNs in the redshift range between 0.3 and 5
by applying a Bayesian approach \citep{kelly09}. Hereafter we label
BHMF and ERDF derived with the Bayesian approach with {\it estimated}.
The {\it estimated} BHMF of broad-line AGNs shows an increase in
number density above $M_{\rm BH} = 1\times10^9 M_{\sun}$ 
from $z=0$ to 2. In contrast, lower mass SMBHs show a relatively
flat number density evolution up to $z=2$. The ''down-sizing" trend
expected from the AGN luminosity function is confirmed
by the steeper decrease of active SMBHs in the higher mass range
from $z=2$ to $0$. However, it needs to be noted
that broad-line AGNs in the SDSS sample only cover large $M_{\rm BH}$
and large $\lambda_{\rm Edd}$ AGNs in the redshift range. For example, at 
$z=1.5$ the sample is only 30\% complete down to $M_{\rm BH}\sim 10^{9} M_{\sun}$
and $\lambda_{\rm Edd}\sim0.6$ \citep{kelly10}. Due to the shallow
detection limit, 
the discrepancy between the {\it binned} and {\it estimated} broad-line
AGN BHMFs is as 
large as two orders of magnitude in the mass range around 
$M_{\rm BH}=10^{9} M_{\sun}$ at $z=1.4$ \citep{shen12}. Therefore, in 
order to examine the cosmological evolution of BHMFs and ERDFs, 
a sample of broad-line AGNs with fainter detection limits is 
needed.

In order to reveal accretion onto SMBHs in an era of violent growth,
we examine the BHMF and ERDF of broad-line AGNs at $z=1.4$
using a sample constructed from the 
X-ray survey of the Subaru {\it XMM-Newton} Deep Survey (SXDS) field.
As suggested by the SMBH growth curves \citep{marconi04}, a significant 
part of the accretion growth of SMBHs is thought to be taking place 
in the redshift range between 1 and 2. Therefore the direct determination of the
BHMF and ERDF in this redshift range 
is of critical importance.
Thanks to the moderately deep detection limit and wide area of the survey, 
we can construct a large sample of broad-line AGNs which is 
one order of magnitude fainter than that available from SDSS. Furthermore, the sample 
covers the flux range around the knee of the X-ray $\log N$-$\log S$ relation
\citep[$1\times10^{-14}$ erg s$^{-1}$ cm$^{-2}$ in the 2--10~keV band;][]{cowie02},
and the sample size is several times larger than the deep {\it Chandra} surveys in this
flux range. The sample represents the population of SMBHs that dominates the accretion
growth of the SMBHs.

This paper is organized as follows.
The details of the sample are described in Section~\ref{sec_SAMPLE}.
Measurements of broad-line width of \ion{Mg}{2} and H$\alpha$ lines and
$M_{\rm BH}$ estimates based on these line widths are 
discussed in Section~\ref{sec_MBH}. In Section~\ref{sec_LE}, estimates 
of the bolometric luminosity and Eddington-ratio are presented. In Section~\ref{sec_FUNC},
we first present the {\it binned} BHMF and ERDF 
calculated applying the $V_{\rm max}$ method to
the sample of broad-line AGNs between $z=1.18$ and $1.68$ 
with black hole mass estimates. 
We then present the
detection-limit corrected broad-line AGN
BHMF and ERDF derived with the Maximum Likelihood
method assuming functional shapes of the BHMF and ERDF. 
In this paper,
we do not include the effect of the uncertainties of the virial black hole mass
estimate in the BHMF and ERDF determination. In Section~\ref{sec_DISCUSS},
the shapes of the {\it corrected} broad-line AGN
BHMF and ERDF are compared with 
those in a  similar redshift range
from SDSS \citep{shen11} and those in the local Universe from the ESO/Hamburg 
survey \citep{schulze10}. The contribution to the {\it binned} active
BHMF from obscured narrow-line AGNs is also discussed.
Throughout this paper we adopt the following cosmological parameters:
$H_0$ = 70 km s$^{-1}$ Mpc$^{-1}$, $\Omega_{\rm M}=0.3$, and
$\Omega_{\Lambda}=0.7$. Magnitudes are given in the 
AB magnitude system \citep{oke74} unless otherwise noted.

\section{SAMPLE}\label{sec_SAMPLE}

\begin{figure}
 \begin{center}
  \includegraphics[scale=0.65]{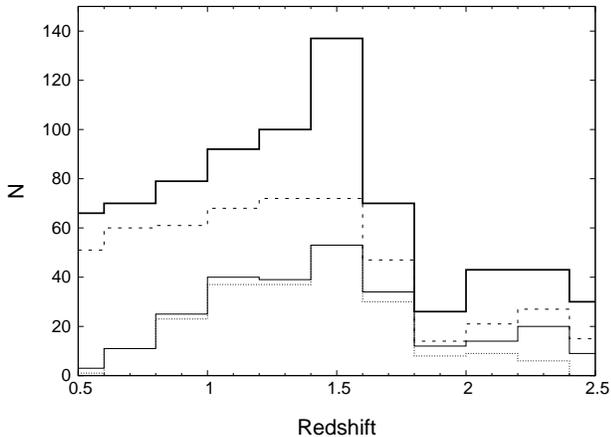}
  \caption{
Redshift distribution of the sample. The thick solid line is
for the entire sample with spectroscopic or photometric redshifts. 
AGNs with spectroscopic redshifts are shown with short-dashed line. 
The thin solid line is the broad-line AGNs with spectroscopic redshifts.
The dotted line is the broad-line AGNs with a black hole mass estimate
from either the \ion{Mg}{2} or H$\alpha$ broad-line. 
\label{fig_red}}
 \end{center}
\end{figure}

\begin{figure}
 \begin{center}
  \includegraphics[scale=0.85]{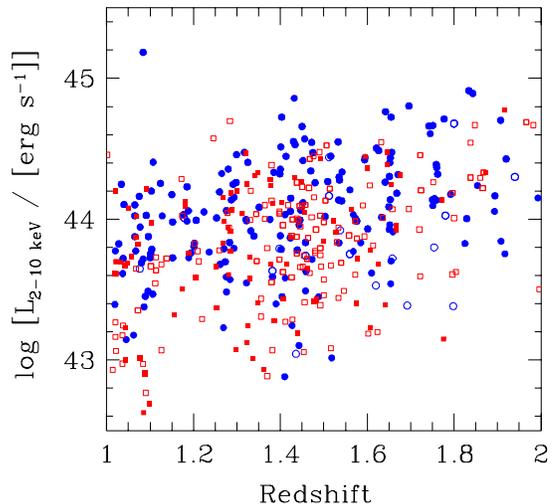}
  \caption{
Redshift vs. absorption-corrected 2--10~keV luminosity
of X-ray-selected AGNs in the SXDS. {\it Filled circles} and {\it squares}
represent broad-line and narrow-line AGNs with spectroscopic 
identification, respectively.
Broad-line (narrow-line) AGN candidates with photometric redshifts are
shown with {\it open circles} ({\it open squares}).
\label{SXDS_red_lum_14}}
 \end{center}
\end{figure}

\begin{deluxetable*}{lrrrl}
\tabletypesize{\footnotesize}
\tablecaption{Summary of the SXDS broad-line AGN sample \label{tbl_sample}}
\tablewidth{0pt}
\tablehead{
  \colhead{ } & \colhead{N} & \colhead{${\rm{N}}_{\rm{soft}}$} &
  \colhead{${\rm{N}}_{\rm{hard}}$} & \colhead{Note}
}
\startdata
X-ray sources                        & 945 & 781 & 584 & within deep Suprime-cam image coverage \\
AGN candidates                       & 896 & 733 & 576 & \\
Optical spec. observed               & 590 & 517 & 396 & among the 896 AGN candidates \\
FMOS spec. observed                  & 851 & 704 & 548 & among the 896 AGN candidates \\
Spec. identified                     & 586 & 514 & 397 & \\
\ion{Mg}{2} broad-line               & 186 & 181 & 137 & z range 0.489 - 2.329 \\
H$\alpha$ broad-line                 &  81 &  78 &  68 & z range 0.634 - 1.655 \\
\ion{Mg}{2} and H$\alpha$ broad-line &  52 &  51 &  44 & \\
\enddata
\tablecomments{}
\end{deluxetable*}

\begin{deluxetable*}{lrrr}
\tabletypesize{\footnotesize}
\tablecaption{Summary of SXDS AGNs in the redshift range $1.18 \leq {\rm{z}} \leq 1.68$\label{tbl_MFsample}}
\tablewidth{0pt}
\tablehead{
  \colhead{ } & \colhead{N} & \colhead{${\rm{N}}_{\rm{soft}}$} &
  \colhead{${\rm{N}}_{\rm{hard}}$}
}
\startdata
Broad-line AGN sample                                                                       &     &     &     \\ \hline
Total (excluding broad-line AGNs with photometric redshift) \tablenotemark{a}               & 118 & 112 &  90 \\
$M_{\rm BH}$ with \ion{Mg}{2} broad-line FWHM                                               &  93 &  90 &  67 \\
Additional $M_{\rm BH}$ with H$\alpha$ broad-line FWHM                                      &  23 &  21 &  21 \\
Spectroscopically-identified broad-line AGN w/o \ion{Mg}{2} or H$\alpha$ measurement        &   2 &   1 &   2 \\
Broad-line AGN in the redshift range identified with photometric redshift \tablenotemark{a} &  10 &  10 &   5 \\ 
                                                                                            &     &     &     \\
Narrow-line AGN sample                                                                      &     &     &     \\ \hline
Total                                                                                       & 158 & 120 & 101 \\
Spectroscopically-identified narrow-line AGN                                                &  66 &  52 &  41 \\
Narrow-line AGN with photometric redshift                                                   &  92 &  68 &  60 \\
\enddata
\tablenotetext{a}{Non-detection of broad emission line in the observed wavelength range implies that
most of the broad-line AGNs with photometric redshift may not be in the redshift range.
Therefore we do not include them from the total number in the first line and the derivation of
BHMF and ERDF, but show their number explicitly in the 5th line for reference. 
See text in Section 2.} 
\end{deluxetable*}

A sample of $z\sim1.4$ AGNs is constructed from X-ray observations of
the SXDS field \citep{ueda08}. The field was observed with the {\it XMM-Newton} covering
one central $30^{\prime}$ diameter field at a depth of 100ks exposure and six flanking fields
with 50ks exposure time each \citep{ueda08}. 
From summed images of pn, MOS1, and MOS2
detectors, there are 866 and 645 sources detected in the 
0.5--2~keV (soft) and 2--10~keV (hard) bands, respectively, with a 
detection likelihood, which is determined by 
point spread function fitting, larger 
than 7, which corresponds to a confidence level of 99.9\%. In this paper, 
only X-ray sources in the region covered with deep optical imaging data taken with 
Suprime-cam on the Subaru telescope \citep{furusawa08} are considered. 
There are 781 and 584 sources in the soft- and hard-bands, respectively. 
Once Galactic stars and clusters of galaxies candidates are
removed, 733 (576) sources remain as AGN candidates. Hereafter we call the 
former (latter) sample the soft- (hard-) band sample. The detection limit of the
survey corresponds to a flux of $6\times10^{-16}$ erg s$^{-1}$ cm$^{-2}$ 
($3\times10^{-15}$ erg s$^{-1}$ cm$^{-2}$) in the soft (hard) band.
The area covered with the flux limit is 0.05 deg$^{2}$, and 1.0 deg$^{2}$
is covered at the flux limit of $4\times10^{-15}$ erg s$^{-1}$ cm$^{-2}$ 
in the soft band.
Considering sources common to both samples, 
there are 896 unique AGN candidates in total (see Table~\ref{tbl_sample}).

In order to identify the X-ray sources spectroscopically, optical
observations were conducted with various multi-object
spectrographs on 4m- and 8m-class telescopes. The optical spectroscopic
observations cover 590 out of the 896 sources in total. 
Even though the observations do not cover the entire sample, they are 
not heavily biased toward a specific type of object, since we do not use 
any further discrimination such as color in the target selection in most
of the observations. Details of the observations
are summarized in Akiyama et al. (2012, in preparation).

Additional intensive NIR
spectroscopic observations were made with the Fiber Multi Object Spectrograph
(FMOS) on the Subaru telescope \citep{kimura10}. This instrument can observe up to 200 objects 
simultaneously over a
30$^{\prime}$ diameter FoV in the cross-beam switching mode with two spectrographs. 
The spectrographs cover the wavelength range between 9000{\AA} and 18000{\AA}
with spectral resolution of $R\sim800$ at $\lambda\sim1.55 \mu$m in the low-resolution mode. 
A total of 851 sources were observed with this setup during guaranteed, engineering 
and open-use (S11B-098 Silverman et al. and S11B-048 Yabe et al.) 
observations.
The optical and NIR observations spectroscopically identify 586 out of the 896
sources. The optical and NIR spectra obtained in the identification 
observations are used for the broad-line width measurements
described in the next section.

The remaining 310 sources cannot be identified spectroscopically,
mostly because of their faintness. Most of them are fainter than 
$i=23.5$ magnitude. For such objects, photometric redshifts have been 
estimated using the HyperZ photometric redshift code \citep{bolzonella00} with galaxy 
and QSO Spectral Energy Distribution (SED) templates. Photometric data 
in 15 bands covering from 1500{\AA} to 8.0$\mu$m are
used in the estimation. 
In order to reduce the number of AGNs with significantly
different photometric redshift from spectroscopic one
("outliers"), we apply two additional constraints in addition
to the $\chi^2$ minimization considering the properties of the
spectroscopically-identified AGNs. First one is that the objects
with stellar morphology in the deep optical images are $z>1$
broad-line AGNs. Almost all X-ray sources with stellar morphology
are identified with broad-line AGNs at $z>1$ in SXDS. They show
a bright nucleus and their observed optical light is dominated
by the nuclear component. Second one is the
absolute magnitude range of the galaxy and QSO templates.
Considering the absolute magnitude range of
spectroscopically-identified broad-line and narrow-line AGNs,
we limit the $z$-band absolute magnitude range of the
galaxy (QSO) template between $M_{z}=-20.0--25.0$
(mag) ($M_{z}=-22.0--26.5$ (mag)). 

The accuracy of the photometric redshifts are examined by
comparing them with the spectroscopic redshifts. The median
of $\Delta z / (1+z_{\rm spec})$ is 0.06 for the entire
sample. We further examine the accuracy by the normalized
median absolute deviation (NMAD; $\sigma_{z}$)
following \citet{brammer08}. For the entire sample,
$\sigma_{z}$ is 0.104, which is larger than that of
the photometric redshift estimations for X-ray-selected
AGNs with medium band filters \citep{cardamone10, luo10}. 
The $\sigma_{z}$ for broad-line AGNs
(0.201) is larger than that for narrow-line AGNs (0.095).
This is because there is no strong feature in the SEDs
of the broad-line AGNs except for the break below Ly$\alpha$.

From the photometric redshift determination, not only their photometric
redshifts, but also their types of SED can be constrained. For spectroscopically
identified AGNs, there is a good correlation between the spectral type and
the best-fit SED type; narrow-line and broad-line AGNs are fitted well with 
galaxy and QSO templates, respectively. Therefore, we classify objects
fitted better with the QSO templates as broad-line AGNs and the others
as narrow-line AGNs. The classification does not perfectly
match the spectroscopic classification: in the redshift range, 10 out of 66
(31 out of 118)
spectroscopically-identified narrow-line (broad-line) AGNs are photometrically
classified as broad-line (narrow-line) AGNs.
The SEDs of the
spectroscopically-unidentified objects suggest that most of them are
obscured narrow-line AGNs above redshift 1. For 6 objects, no photometric
redshift can be estimated because they are detected only in a few bands.
They are faint and are unlikely to be broad-line AGNs
in the redshift range between $1.18$ and $1.68$.
Further details of the photometric redshift determination is
discussed in Akiyama et al. (2012, in preparation).

In this paper, for objects with spectroscopic identification,
we designate objects as broad-line AGNs if they show 
either \ion{Mg}{2}$\lambda\lambda 2796,2803$ 
or H$\alpha$ emission lines with width greater than 1000 km s$^{-1}$.
We estimate black hole mass of the broad-line AGNs with
either broad \ion{Mg}{2} or H$\alpha$ line.
The threshold is narrower than typical threshold 
used to discriminate broad-line AGNs (1500 or 2000 km s$^{-1}$). 
We determine the threshold
considering the distribution of the FWHM of the broad-line AGNs
in the local universe \citep{hao05, stern12}. The broad-line AGNs with
the line FWHM close to the threshold correspond to the narrow-line Seyfert 1s.
Broad \ion{Mg}{2} and H$\alpha$ lines are detected for 186 and 81 AGNs respectively, 
with redshifts in the range between $0.5$ and $2.3$.
For 52 objects, both broad \ion{Mg}{2} and H$\alpha$ lines are detected. 
For 29 objects, a broad-line is  only detected in H$\alpha$. This is mostly
due to the lack of optical spectra. Only 4 AGNs
(SXDS0215, SXDS0387, SXDS0527, SXDS0728) with broad
H$\alpha$ line show no broad \ion{Mg}{2} line, although their optical spectra
cover the \ion{Mg}{2} wavelength region and are deep enough to detect
continuum emission. Such AGNs are thought to
be moderately affected by dust extinction and we correct for this in the
determination of their continuum luminosity for the black hole
mass and the intrinsic bolometric luminosity estimates.  
Details are given in Section~\ref{sec_MBH} and \ref{sec_LE}.

For the derivation of the broad-line AGN BHMF,
we limit the sample to the redshift range
between 1.18 and 1.68. The FMOS $H$-band observation covers the broad H$\alpha$
line for AGNs in the redshift range. Considering the typical detection
limit for the broad H$\alpha$ line in this redshift range
($L_{\rm H\alpha}=2\times10^{42}$ (erg s$^{-1}$)), we expect a broad H$\alpha$
can be detected for broad-line AGNs brighter than 
$L_{\rm 2-10~keV}=3\times10^{43}$ (erg s$^{-1}$), if they have
the H$\alpha$ to hard X-ray luminosity ratio typical of 
broad-line AGNs \citep{ward88}. Photometric redshift
estimates suggest 10 of the unidentified sources can be
broad-line AGNs in this redshift range. All but one
of the 10 candidates have estimated hard X-ray luminosity larger
than $3\times10^{43}$ (erg s$^{-1}$). But, 8 of the 10
broad-line AGN candidates do not show a broad-line in 
the FMOS observations. Considering the
uncertainty of photometric redshift for broad-line AGNs,
we expect they are broad-line AGNs at outside of the
redshift range. However,
there is a 0.4 dex scatter between the $L_{\rm 2-10~keV}$
and $L_{\rm H\alpha}$\citep{ward88}, and it is still possible that
they have weaker broad H$\alpha$ line than the typical
broad-line AGNs. Furthermore, additional one broad-line 
AGN candidate does not have broad-line in the optical
spectroscopy, it may also be a broad-line
AGN at outside of the redshift range. Considering the
non-detection of broad-line in the observed wavelength
range, we do not include the 10 photometric candidates of 
broad-line AGNs in the redshift range in the derivation of
the BHMF and ERDF below.
The numbers of X-ray-selected AGNs in the redshift range are
summarized in Table~\ref{tbl_MFsample}. The median 
redshift of the sample is 1.43.

The redshift distribution of the sample is shown in Figure~\ref{fig_red}
for the redshift range  0.5 to 2.5.
The thick solid line shows the redshift distribution of the 
X-ray AGNs with spectroscopic or photometric redshifts. The dashed line
shows the distribution of spectroscopically-identified AGNs.
The thin solid line shows the distribution for broad-line AGNs with spectroscopic redshifts.
The dotted line is the distribution for broad-line AGNs that have
black hole mass estimates with either broad \ion{Mg}{2} or H$\alpha$ emission lines. 

In Figure~\ref{SXDS_red_lum_14}, the absorption-corrected 2--10~keV luminosity of
AGNs are shown as a function of redshift. Following \citet{ueda03},
we assume the intrinsic shape of the X-ray spectrum and estimate the
intrinsic column density from the observed hardness ratio and redshift.
The intrinsic X-ray spectrum of AGNs is modeled by a combination of
a power-law component with 
high-energy cut off  $E^{-\Gamma} \times \exp{(-E/E_{c})}$ and
a reflection component. A photon index
$\Gamma$ of 1.9 and cutoff energy $E_{c}$ of 300~keV are assumed.
We calculate the reflection component with the  ``pexrav'' \citep{magdziarz95}
model in the XSPEC package assuming a solid angle of 2$\pi$, an inclination angle
of $\cos i = 0.5$, and solar abundance of all elements.
Strength of the reflection component is about 10\% of the direct
component just below 7.1~keV.
The intrinsic SEDs are modified with intrinsic photo-electric absorption
described by a hydrogen column density, $N_{\rm H}$.
The $N_{\rm H}$ value of each object is evaluated from the observed
0.5--2~keV and 2--4.5~keV hardness ratio.
The absorption-corrected luminosity in the 2--10~keV band is derived 
from the observed 2--10~keV count-rate by
correcting for the photo-electric absorption.
For objects only detected in the 0.5--2~keV band, the count-rate
in that band is used instead of the 2--10~keV count-rate.
Most of the broad-line AGNs have a hardness ratio consistent
with no significant absorption and the required correction is small. 
Only 5 out of the 215 broad-line AGNs have $\log N_{\rm H}$ as 
large as $23$; for these the correction in luminosity is as large as 0.5 dex.

The absorption-corrected 2--10~keV luminosity of AGNs distribute
between $L_{\rm 2-10~keV}=10^{43}$ and $10^{45}$ (erg s$^{-1}$) in the
redshift range of $z=1-2$. The SXDS AGNs cover the most important part
of the accretion growth of the SMBHs. AGNs in the luminosity range dominate
the hard X-ray luminosity density of the Universe in the redshift range, 
furthermore, the hard X-ray luminosity density as a function of redshift
peaks at $z=1$ \citep{aird10}.

\section{LINE WIDTH AND LUMINOSITY MEASUREMENT}\label{sec_MBH}

\subsection{Method for Black Hole Mass Estimation}

Assuming that the scaling relationship between
luminosity and broad-line region size 
derived by reverberation mapping for local broad-line AGNs
is applicable to broad-line AGNs at high-redshifts,
black hole masses of broad-line AGNs can be estimated from
their continuum luminosities and line widths of the broad lines.
We estimate the black hole mass of $z\sim1.4$ broad-line AGNs with
\ion{Mg}{2} broad-line widths measured in the optical spectra.
As the optical spectroscopy does not cover the entire sample,
we supplement these with H$\alpha$ broad-line widths measured
in the NIR spectra. In this subsection, we first introduce the 
equation used to estimate the black hole mass.

There are several calibrations available for the black
hole mass estimation using the  \ion{Mg}{2} broad line 
\citep{mclure02, mclure04, mcgill08, vestergaard09, wang09, shen11, rafiee11}.
We use the black hole mass estimate from the \ion{Mg}{2} 
broad-line FWHM calibrated by broad-line QSOs in the SDSS DR3 
\citep{vestergaard09}. The estimation is consistent to
within 0.1 dex of the H$\beta$ and \ion{C}{4} 
mass estimates with single-epoch spectra. They
are calibrated to the black hole mass from the 
reverberation mapping \citep{vestergaard06}. 
It needs to be noted that the black 
hole mass determined with the single-epoch H$\beta$
spectrum typically has scatter of 0.4-0.5 dex around that
from the reverberation mapping \citep{vestergaard06}.
\citet{park12} also estimate the uncertainty of the 
single-epoch black hole mass to be 0.4-0.5 dex.
Furthermore, if AGNs are
close to the Eddington limit, the influence
of radiation pressure may cause an underestimation of the 
black hole mass by $0.5$ dex \citep{marconi08}.

Because the 3000{\AA} monochromatic luminosity
is available for most of the broad-line AGNs with spectroscopic 
observations, we use the following equation from \citet{vestergaard09}
incorporating the 3000{\AA} monochromatic luminosity, $L_{\lambda 3000}$. 
\begin{eqnarray}
M_{\rm BH} [M_{\sun}] \nonumber \\
&=& 10^{6.86} \left[ \frac{\rm FWHM_{\rm MgII}}{1000 {\hspace{1mm}} \rm{km \ s^{-1}}}\right]^2
  \left[ \frac{\lambda 3000 L_{\lambda 3000} 
  } {10^{44}{\hspace{1mm}}{\rm{erg \ s^{-1}}}} \right]^{0.5}.
\label{eq_MBH}
\end{eqnarray}
In this calibration, the FWHM of broad \ion{Mg}{2} line,
FWHM$_{\rm MgII}$, is measured 
by fitting multiple Gaussians to the \ion{Mg}{2} emission line profile. 
\citet{vestergaard09} remove narrow-line 
component of \ion{Mg}{2} if necessary \citep{vestergaard11}.
In other calibrations, such
as \citet{mclure02} and \citet{mclure04}, fit the \ion{Mg}{2}
profile with a single broad-line and narrow-line components.
\citet{rafiee11} use the line dispersion, $\sigma$, of the broad-line
because $\sigma$ correlates more tightly with the delay observed in 
reverberation mapping, i.e. size of the broad-line region, 
 than with the FWHM \citep{peterson04}. 

In Equation (1), the broad-line region size $R$
is assumed to follow $L^{\alpha}$ with $\alpha$ of 0.5,
which is equivalent to assuming that the broad-line regions 
of various AGNs can be described as having similar ionization states, 
ionizing photon spectra and electron densities. 
The value of
$\alpha$, based on reverberation mapping results for H$\beta$ broad-lines, is 
determined to be 0.47, $0.62\pm0.14$, and $0.518\pm0.039$ \citep{mclure02, mclure04, bentz06}, respectively, all
consistent with $\alpha$ of 0.5 within the uncertainties.

Regarding the coefficient for the virial product
($\epsilon$ of $M_{\rm BH}=\epsilon {\rm FWHM_{H\beta}}^2 L^{0.5}$
or $f$ of $M_{\rm BH}=f \sigma_{H\beta}^{2} L^{0.5}$),
Equation (1) is based on the calibration done by 
\citet{onken04} ($\epsilon$ of 1.4 or $f$ of 5.5)
assuming the estimated $M_{\rm BH}$ of local broad-line AGNs 
from reverberation mapping and their bulge velocity dispersions
follow the black hole mass and the bulge velocity dispersion
relation of non-active galaxies in the local Universe.
The coefficient is consistent with that
derived with local Seyfert 1 galaxies \citep[$f=5.2$;][]{woo10}.
Recent calibration shows broad-line AGNs hosted in barred galaxies are
consistent with significantly smaller values \citep[$f=2.3$;][]{graham11},
but the range of black hole mass of the barred galaxies is
smaller than the current sample, and non-barred galaxies with
larger $M_{\rm BH}$ is consistent with $f\sim5.4$ \citep{graham11},
thus we use Equation (1).

The optical spectroscopic observations do not cover all of the
broad-line AGNs at $z=1.18-1.68$; 25 out of 118 spectroscopically-identified 
broad-line AGNs in the redshift range do not have \ion{Mg}{2} data.
Therefore, we also utilize the H$\alpha$ FWHM in addition to the \ion{Mg}{2} FWHM; 
additional 23 broad-line AGNs have H$\alpha$ broad-line data. 
Although the FMOS spectra cover H$\beta$ in $J$-band,
the strength of the H$\beta$ broad-line is 3 times or more weaker than
the H$\alpha$ broad-line and the uncertainty of the FWHM of the broad H$\beta$
is significantly larger than that of the broad H$\alpha$ line. 
Therefore, we do not use the H$\beta$ FWHM.
Because ionization potentials of hydrogen and
\ion{Mg}{2} are similar, Balmer and \ion{Mg}{2}$\lambda\lambda 2796,2803$
broad lines are expected to be emitted in a similar region. A detailed
photoionization model calculation indicates that the
equivalent widths of H$\alpha$ and \ion{Mg}{2} lines 
have a similar dependency on the cloud density and 
ionization parameter, i.e., they are emitted from similar broad-line
clouds \citep{korista97}. Considering this similarity, we use
the same black hole mass equation employed for \ion{Mg}{2} FWHM above for
H$\alpha$ FWHM, after correcting for a small systematic
difference between \ion{Mg}{2} FWHM and H$\alpha$ FWHM, as detailed below.
We do not use the scaling relation calibrated for 
H$\alpha$ broad-line \citep{greene05} in order to be
consistent within our sample. 
The derivation of the 3000{\AA} monochromatic luminosity is discussed
in Section~\ref{sec_UVOpt}.

\subsection{\ion{Mg}{2} Line Width Measurements with Optical Spectrum}\label{sec_Mg2}

\begin{figure*}
 \begin{center}
  \includegraphics[scale=0.65]{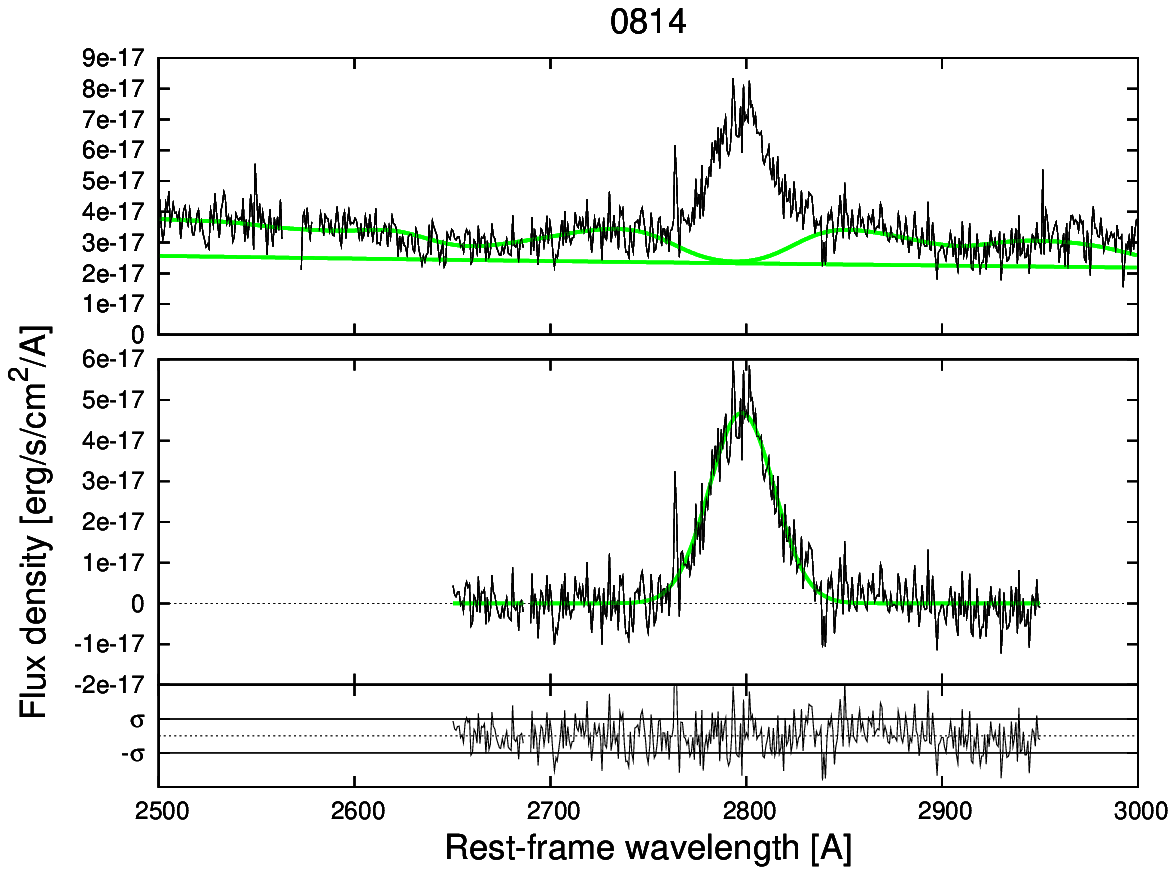}
  \includegraphics[scale=0.65]{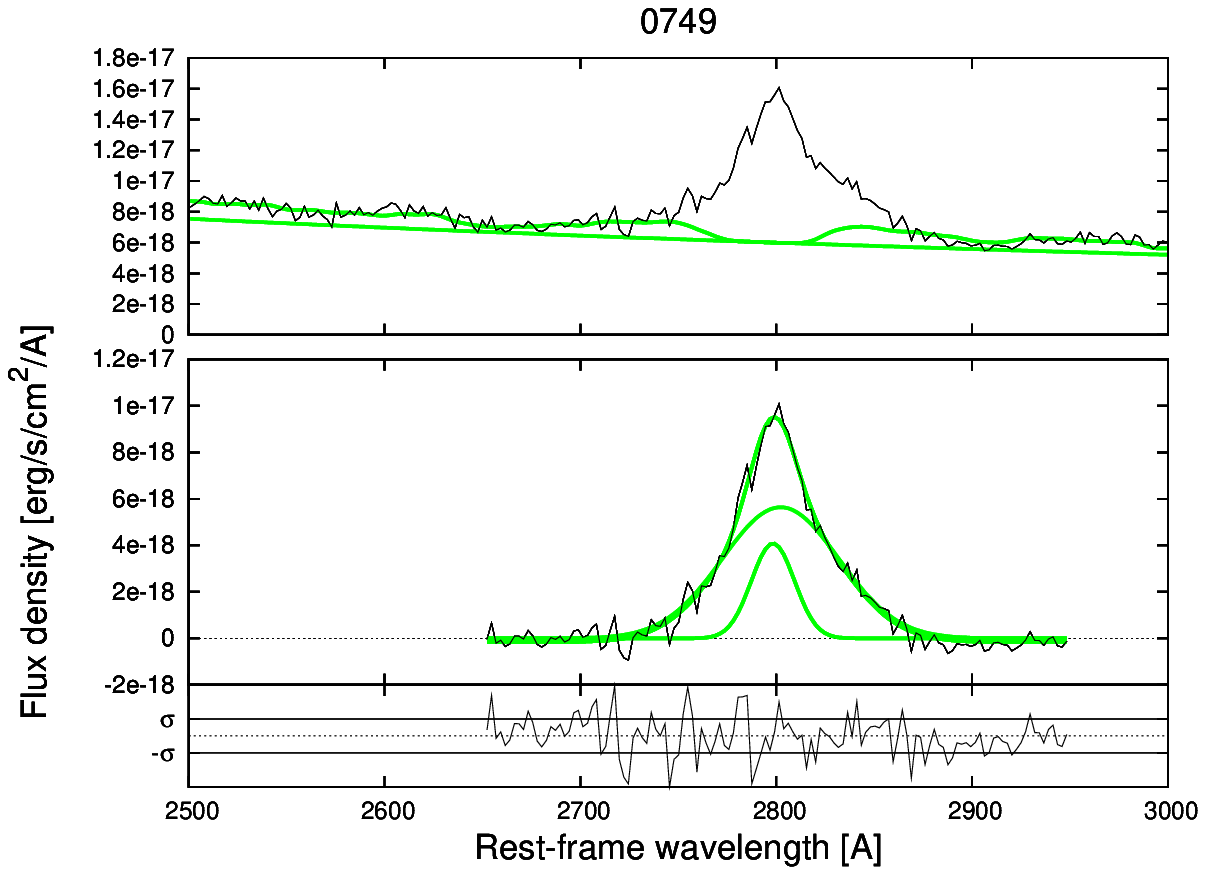}
  \caption{
Examples of broad \ion{Mg}{2} fitting. Upper panels
show the observed data ({\it thin solid line}) and the
best fit model from the power-law continuum and \ion{Fe}{2}
fitting ({\it thick solid line}). Power-law component
of the best fit model is also shown separately.
Middle panels show the results of the broad \ion{Mg}{2} line
fitting. Pure \ion{Mg}{2} component after subtracting the 
power-law continuum and \ion{Fe}{2} components and best
fit model from the \ion{Mg}{2} fitting are
shown with {\it thin solid line} and {\it thick solid line},
respectively.
Each component of the best fit model is also shown.
Only the wavelength range used for the broad-line
fitting is plotted.
Bottom panels show the residual after fitting.
Left) SXDS0814 with single broad-line, and
right) SXDS0749 with 2 components.
\label{Mg2_fitting}}
 \end{center}
\end{figure*}

\begin{figure}
 \begin{center}
  \includegraphics[scale=0.8]{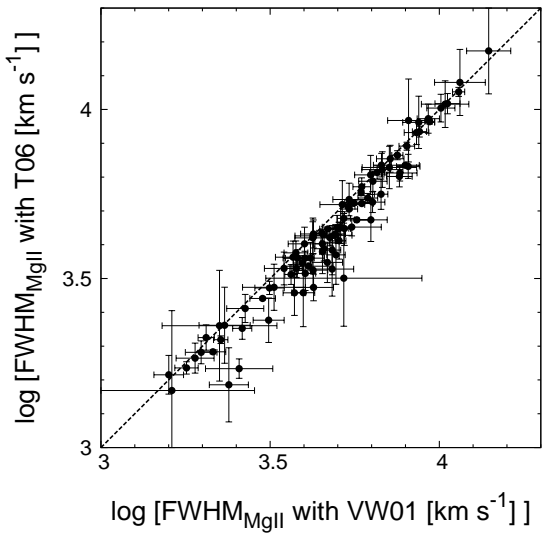}
  \caption{
FWHMs of the broad \ion{Mg}{2} emission line determined
after \ion{Fe}{2} fitting with
with VW01 and T06 \ion{Fe}{2} templates. The dashed line shows
the equal FWHM line.
\label{fig_Fe_com}}
 \end{center}
\end{figure}

Optical spectra of the AGNs were obtained with various 
instruments on 4-8m class telescopes such as 2dF on the Anglo-Australian 
Telescope, VIMOS and FORS on the Very Large Telescope, FOCAS on the Subaru telescope,
DEIMOS on the Keck telescope, and IMACS on the Magellan telescope. 
Most of them were obtained with spectral resolution of $R\sim250-500$.
Details of the observations are described in Akiyama et al. (2012, in preparation).
The optical spectroscopic data were reduced using standard procedures
and are corrected for the
dependence of the sensitivity on wavelength with standard star observations. 
We further correct the normalization of the spectra to match the
observed $R$-band magnitudes in the deep Suprime-cam images.
The optical photometric data are corrected
for the Galactic extinction in the SXDS field ($A_{R}$ of 0.07 mag).
We do not correct the optical spectroscopic data for the wavelength
dependence of the Galactic extinction which is negligibly small.
This correction does not affect the line width measurement, but
affects the continuum flux measurement. The normalization can be 
affected by the variability of broad-line AGNs between the epochs of 
the imaging and the spectroscopic observations. Typically there is one year
gap between the imaging and spectroscopic observations. The structure
function of optical variability of QSOs \citep{cristiani96} suggests that 
there can be 0.2 mag variation during the time lag on average. Therefore, the 
uncertainty of $M_{\rm BH}$
due to the time-variation of AGNs is expected to be 0.04 dex.

In order to determine the \ion{Mg}{2} FWHMs of 
broad-line AGNs, it is necessary to consider \ion{Fe}{2} emission 
lines as well as the power-law continuum component 
in the UV wavelength range. Because there are
many broad \ion{Fe}{2} emission lines in the wavelength range, 
they look like an additional continuum component to the
power-law continuum of broad-line AGNs. We use a \ion{Fe}{2} template
derived from the UV spectrum of the narrow-line Seyfert 1 
galaxy, I Zw 1 \citep{vestergaard01}. This template
covers the rest-frame wavelength range between 
1074{\AA} and 3089{\AA}. We do not include the Balmer
continuum in the fitting, because the
wavelength coverage is not wide enough to constrain 
its contribution. The ignorance of the Balmer continuum
does not affect the \ion{Mg}{2} FWHM measurements
significantly, but the luminosity of the power-law
continuum can be overestimated by 0.12 dex \citep{shen12b}.

A fit to the power-law continuum, \ion{Fe}{2} emission lines, and
\ion{Mg}{2} emission line is carried out as follows. 
First, we determine the normalization of \ion{Fe}{2} 
and continuum component using $\chi^2$ minimization
in the two rest-frame wavelength ranges, 2500-2700{\AA}
and 2900-3000{\AA} in which \ion{Mg}{2} broad-line component
is negligible. These are close to the pure Fe emission windows
nos.9 and 10 in \citet{vestergaard01}. We vary the line width 
of the \ion{Fe}{2} emission lines from 1000 km s$^{-1}$ to 
15000 km s$^{-1}$ with a step size of 250 km s$^{-1}$ by convolving
a Gaussian profile with the \ion{Fe}{2} template which has a velocity
width of 900 km s$^{-1}$. We assume a constant line width for
all \ion{Fe}{2} emission lines. The scaling of the \ion{Fe}{2} emission 
is changed from 0 to 100\% of the observed continuum level
with a step size of 0.01\%. The continuum component is modeled
with a power-law spectrum ($f_{\nu} \propto \nu^{-\alpha}$). 
The observed wavelength ranges affected by strong night sky 
lines (5555-5605{\AA} and 6270-6320{\AA})
are removed in the fitting. 
Because some optical spectra do not have
noise level estimation from the standard reduction method,
we estimate the noise level for each spectrum as follows.
Considering that the noise level does not vary significantly 
within the observed wavelength range, we use a constant noise level
for the entire wavelength range. In the first stage of the fitting, we use 
the standard deviation determined within the wavelength
range as the noise level. Subsequently, we determine 
the noise level from the rms of the residual of the first fitting,
and carry out the final fit for the continuum. The noise level 
is used for the \ion{Mg}{2} line profile fitting as well.
Examples of continuum fits are shown in the upper panels of
Figure~\ref{Mg2_fitting}.

By subtracting the \ion{Fe}{2} emission and power-law continuum 
components, the \ion{Mg}{2} broad-line component is extracted. 
Then we measure the FWHM of the \ion{Mg}{2} broad-emission line 
after fitting its line profile with multiple Gaussians using $\chi^2$ minimization. 
In the
fitting, we do not consider the doublet component of
\ion{Mg}{2}$\lambda\lambda 2796,2803$, because the separation is
small and does not affect the measured width of the 
broad \ion{Mg}{2} line. We use mpfit package for python for $\chi^2$ minimization
\citep{markwardt09}. 
Mpfit uses Levenberg-marquardt algorithm to 
derive the best fit parameters. For some objects, the \ion{Mg}{2} 
line profile cannot be described with a single Gaussian component.
In such cases, we consider up to three broad Gaussians
for the broad line. If necessary, we include narrow doublet absorption
lines. Sometimes we also include a narrow Gaussian component
to remove artificial spiky noise features. 
Because no object shows a significant existence of
narrow \ion{Mg}{2} line, we do not include a narrow-emission
line component in the fitting. No inclusion of the narrow-line
component differs from the fitting method used in most of the
literatures such as \citet{mclure02}.
Once the pure \ion{Mg}{2} broad-line component is fitted with the
multiple Gaussian components, the FWHM of the broad \ion{Mg}{2} line is
measured with the best fit profile constructed by combining the
multiple Gaussians. We do not include absorption lines in the combination.
We introduce multiple Gaussian components in order to reproduce the
observed \ion{Mg}{2} broad-line profile smoothly and here we are not
concerned with the physical meaning of the difference from the single Gaussian profile.
Examples of the resulting \ion{Mg}{2} broad-line fitting are shown in
the middle and bottom panels of Figure~\ref{Mg2_fitting}.
Because the spectra are obtained with various instruments, 
the spectral resolution of data varies from object to object. The spectral resolution
is evaluated for each spectrum by using line width of the arc lamp 
spectrum or night sky emission lines. The measured FWHMs are
corrected for the intrinsic spectral resolution.

The uncertainty of the FWHM values for the combined multiple Gaussian
profile is not available from the fitting with mpfit package
(uncertainty is only available for each Gaussian component).
Therefore we evaluate the uncertainty for the FWHM of each object
from the rms scatter of FWHMs measured in mock spectra 
constructed from the best fit profile of the object. 
We construct the mock spectra as follows. First, the best 
fit multi-Gaussian model is shifted by several pixels in 
wavelength from its original position. Then the shifted
model profile is combined with the residual of the original
fitting. By monotonically increasing the shift 
and changing the sign of residual in each pixel randomly,
we construct 100 mock
spectra. The mock spectra are fitted in the same way as the original
data and FWHMs are measured. Finally the rms scatter of the 
derived FWHMs is used as the uncertainties of the FWHM measurement.
For AGNs whose \ion{Mg}{2} broad-line is fitted with single
Gaussian component, we compare the uncertainties derived 
from the $\chi^{2}$ statistics and from the scatter of the mock 
measurements. They are consistent with each other, 
although the rms scatter of the mock measurements is
slightly smaller than the uncertainties from the $\chi^{2}$ statistics.
Hereafter, we use the uncertainty derived with 
the scatter of the mock measurements. The fitting results
are summarized in Table~\ref{tbl_FitRes}. 
Column 4 of the table describes the model used for each object;
"OneBL", "TwoBL" and "ThreeBL" indicate fitted with one, two and three 
broad lines respectively. "OneBLOneAbs" indicates a model with one broad and one
absorption line. 

For the \ion{Mg}{2} profile fitting, we also use the specfit software
in the {\bf stsdas} package of IRAF. This package uses Marquardt algorithm
or simplex algorithm for $\chi^2$ minimization.
We compare the results obtained with mpfit and specfit for
each object. The rms scatter of the difference in FWHMs 
from the two measurements is 0.06 dex. The resulting uncertainty 
in $M_{\rm BH}$ due to the scatter is 0.12 dex. We use the results
obtained with mpfit hereafter. 

The measured FWHM of \ion{Mg}{2} broad-line can be affected by 
the template used for the \ion{Fe}{2} fitting.
For the \ion{Fe}{2} fitting, we use the empirically derived \ion{Fe}{2} template
from \citet{vestergaard01} (VW01 template). There is no \ion{Fe}{2} emission in
the wavelength range between 2770{\AA} and 2820{\AA} in the template. 
\citet{tsuzuki06} also derived the \ion{Fe}{2} template (T06 template) from
UV and optical spectra of I Zw 1. Utilizing the wide wavelength 
coverage available, they fit the continuum with a power-law and Balmer 
continuum and also utilize the H$\alpha$ line profile
for removing the \ion{Mg}{2} line component. The important difference
between the two templates is the \ion{Fe}{2} contribution to the blue wing
of the broad \ion{Mg}{2} line. In the T06 template, the excess wing 
seen on the blue side of the \ion{Mg}{2} line of I Zw 1 compared with 
its H$\alpha$ line profile is regarded as a contribution 
from the \ion{Fe}{2} emission line at around 2790{\AA}.
In order to examine the effect of different \ion{Fe}{2} templates on the FWHM measurement, 
we also apply the \ion{Mg}{2} emission line fitting process described above
using the T06 template for 95 objects 
whose broad \ion{Mg}{2} lines are fitted well with one broad-line
component with the VW01 template. Figure~\ref{fig_Fe_com} compares the $\log$ FWHMs
values derived with the two templates. There is a systematic offset of 0.04 dex;
the FWHMs derived with the T06 template are systematically smaller than those with the VW01 template.
The rms scatter of the difference is 0.05 dex after removing the systematic
offset. The resulting systematic uncertainty of $M_{\rm BH}$ is 0.08 dex. 
In order to compare the broad-line AGN BHMFs
from literature, we follow the same fitting procedure
using the VW01 template, but it needs to be noted
that the FWHMs can have a systematic uncertainty due to the
\ion{Fe}{2} template difference.

\begin{deluxetable*}{crccccccc}
\tabletypesize{\footnotesize}
\tablecaption{Broad-line fitting results for the broad-line AGNs \tablenotemark{a}\label{tbl_FitRes}}
\tablewidth{0pt}
\tablehead{
\colhead{} &
\colhead{} &
\colhead{} &
\multicolumn{2}{c}{\ion{Mg}{2}} &
\multicolumn{2}{c}{H$\alpha$} &
\colhead{} &
\colhead{} \\
\colhead{ID} & 
\colhead{z} & 
\colhead{log[$L_{\rm HX}$} & 
\colhead{FWHM} & 
\colhead{model} & 
\colhead{FWHM} & 
\colhead{model} & 
\colhead{$\log[ \lambda3000 L_{\lambda 3000}$} & 
\colhead{$R-i$} \\ 
\colhead{} & 
\colhead{} & 
\colhead{/ [erg $\rm{s}^{-1}$] ]} & 
\colhead{[km $\rm{s}^{-1}]$} & 
\colhead{} & 
\colhead{[km $\rm{s}^{-1}$]} & 
\colhead{} & 
\colhead{/ [erg $\rm{s}^{-1}$] ]} & 
\colhead{[mag]} 
}
\startdata
0010     & 1.225 & 44.05 & 5341 $\pm$ 12 & TwoBL & 5135 $\pm$ 116 & OneBL & 44.89 & $-$0.04 \\
0018     & 1.452 & 44.42 & \nodata & \nodata & 5488 $\pm$ 92 & BLNL & 45.16 & $-$0.02 \\
0019     & 1.447 & 44.66 & 4823 $\pm$ 312 & OneBL & \nodata & \nodata & 45.41 & $-$0.06 \\
0023     & 1.534 & 44.14 & \nodata & \nodata & 5602 $\pm$ 112 & OneBL & 45.10 & 0.48 \\
0027     & 2.067 & 43.75 & 4136 $\pm$ 1499 & TwoBL & \nodata & \nodata & 45.85 & 0.27 \\
0034     & 0.952 & 43.55 & 4286 $\pm$ 1005 & TwoBL & 2459 $\pm$ 123 & TwoBL & 44.72 & $-$0.15 \\
0036     & 0.884 & 44.16 & 3326 $\pm$ 22 & TwoBL & 2790 $\pm$ 50 & TwoBL & 45.24 & 0.25 \\
0037     & 1.202 & 43.81 & 4516 $\pm$ 70 & TwoBL & \nodata & \nodata & 44.45 & $-$0.06 \\
0050     & 1.411 & 44.03 & 2046 $\pm$ 120 & OneBL & 1800 $\pm$ 68 & OneBL & 44.80 & 0.14 \\
0056     & 1.260 & 44.19 & \nodata & \nodata & 8171 $\pm$ 456 & BLNL & 44.74 & $-$0.03 \\
\enddata
\tablenotetext{a}{Table 3 is published in its entirety in the electronic
edition of ApJ. A portion is shown here for guidance regarding its form
and content.}
\end{deluxetable*}

\subsection{H$\alpha$ Line Width Measurements from NIR Spectrum}

\begin{figure*}
 \begin{center}
  \includegraphics[scale=1.2]{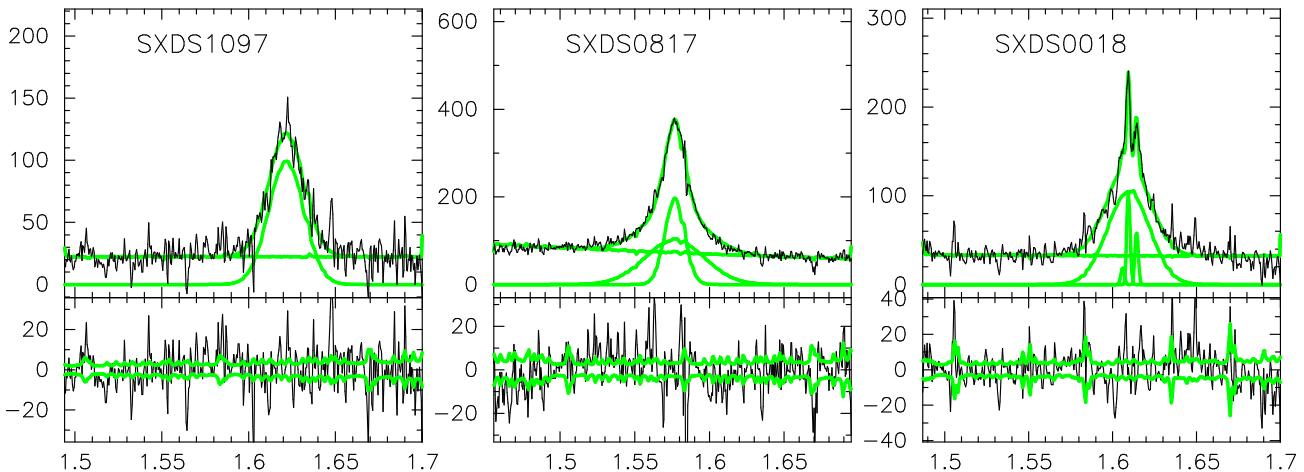}
  \caption{
Examples of broad H$\alpha$ fitting. Upper panels
show the observed data ({\it thin solid line}) and the
best fit model with each component ({\it thick solid lines}). Lower panels
show the residual from the fitting ({\it thin solid line}).
{\it Thick solid lines} in the panel enclose the estimated $1\sigma$ noise
level at each wavelength. Left) SXDS1097 with single
broad-line, middle) SXDS0817 with 2 broad-lines, and
right) SXDS0018 with broad-line and narrow H$\alpha$ and
[\ion{N}{2}]$\lambda\lambda6548,6583$ lines.
\label{Ha_fitting}}
 \end{center}
\end{figure*}

\begin{figure}
 \begin{center}
  \includegraphics[scale=0.9]{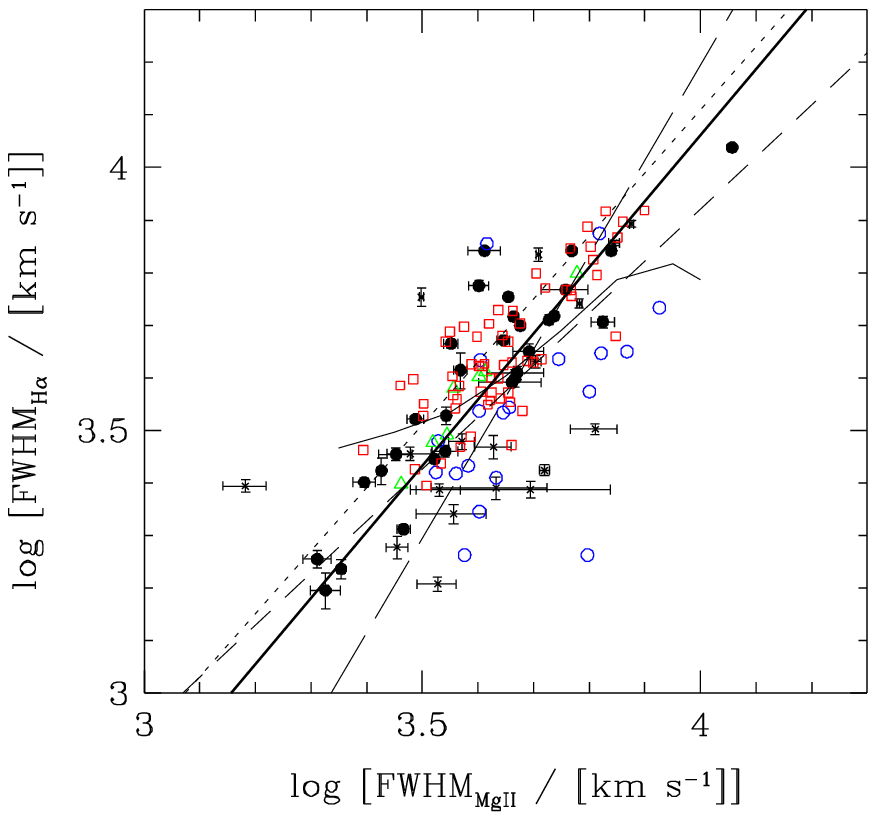}
  \caption{
Comparison between FWHMs of H$\alpha$ and \ion{Mg}{2} broad-lines 
for SXDS AGNs. {\it Filled circles} and {\it crosses} represent 
broad-line AGNs brighter and fainter than $L_{\rm 2-10~keV}=10^{44}$ erg s$^{-1}$,
respectively. The {\it thick solid
line} represents the BCES bisector fitting result for the SXDS AGNs.
{\it Open squares}, {\it triangles}, and {\it circles} are 
AGNs from Shen \& Liu (2012), Greene et al. (2010b), and
McGill et al. (2008), respectively. 
The relationships between the \ion{Mg}{2} and H$\beta$ FWHMs from 
McLure et al. (2002) ({\it short dashed line}), Onken \& Kollmeier (2008) ({\it thin
solid line}), Wang et al. (2009) ({\it dotted line}), and Croom et al. (2011)
({\it long dashed line}) are shown after converting H$\beta$ FWHM to 
H$\alpha$ FWHM.
\label{fig_Mg-H}}
 \end{center}
\end{figure}

The NIR spectra of X-ray sources obtained with FMOS were reduced
with the pipeline data reduction software, FIBRE-pac \citep{iwamuro12}. 
The resulting 1-d spectra are corrected for atmospheric
absorption and sensitivity dependence on wavelength using
relatively bright F-G type stars observed simultaneously with the
targets. Because we do not use the absolute flux of the
continuum component, we do not apply further correction for
normalization based on the photometry. 
Noise level of each
spectrum is also estimated in the pipeline data reduction
software.

The profile of the broad H$\alpha$ line is fitted with 
mpfit in a similar way for broad \ion{Mg}{2} component. The possible existence
of the narrow emission lines of H$\alpha$ and [\ion{N}{2}]$\lambda \lambda$6583,6548 
makes the fitting more complicated than for \ion{Mg}{2}
broad line. We include the
narrow emission lines in the fitting only if a prominent narrow emission
feature, such as asymmetric profile due to the 1:3 flux ratio of [\ion{N}{2}] lines, 
is observed. For the broad H$\alpha$ component, we use
up to two Gaussians to fit their profile. 
In the fitting process, we assume a constant continuum component
because the continuum of the observed NIR spectra does not show 
significant tilt in the region around H$\alpha$ emission line.
Due to the existence of the OH suppression mask, the strength of
the narrow emission line can be underestimated even after the 
sensitivity correction process
if the narrow emission line is close to the masked wavelength.
We apply the fitting procedure taking into account the underestimation
due to the optical masking and the sensitivity correction process.
The details of the fitting procedure are described
in \citet{yabe12}. It needs to be noted that the effect of 
OH suppression mask only affects the narrow-emission line and 
the effects on the broad-emission lines and continuum are negligible.
Examples of H$\alpha$ fitting are shown in Figure~\ref{Ha_fitting}.

The uncertainty in the FWHM measurements of the broad H$\alpha$
emission line is evaluated in the same way as for the broad \ion{Mg}{2} emission lines.
We construct 10 model spectra by adding shifted best fit 
multi-Gaussian model to the residual of the fitting. Then, the same
fitting procedure is applied to the model spectra and the rms scatter
of the measured FWHMs is used as the uncertainty of the FWHM measurement.
In this fitting process we also take into account
the effect of the OH suppression mask.

The resulting FWHM of the broad H$\alpha$ emission is compared with
that of the broad \ion{Mg}{2} line for broad-line AGNs with FWHM measurements
for both lines in Figure~\ref{fig_Mg-H}. We remove \ion{Mg}{2} FWHMs
of 4 broad-line AGNs (SXDS0590, SXDS0630, SXDS0790, and SXDS0969), because the
signal-to-noise ratios of their \ion{Mg}{2} spectra are much lower than those
for their H$\alpha$ spectra. Therefore, 48 broad-line AGNs are plotted in the figure. 
The sample is divided with
the absorption-corrected X-ray luminosity. Broad-line AGNs that are brighter (fainter)
than $L_{\rm 2-10keV}=10^{44}$ erg s$^{-1}$ are marked with {\it filled squares}
({\it crosses}). The measured
FWHMs of the broad \ion{Mg}{2} and H$\alpha$ lines roughly follow the equality line.
However, there may be a tendency that broad-line AGNs with lower
luminosity have systematically smaller broad H$\alpha$ FWHM than broad \ion{Mg}{2} 
line. The distribution is broadly consistent with those of
broad-line AGNs measured in the literature. We show
the \ion{Mg}{2} and H$\alpha$ FWHMs of individual broad-line
AGNs from \citet{shen12b}, \citet{greene10b}, and \citet{mcgill08}.
The samples of \citet{shen12b} and \citet{greene10b} are luminous
broad-line AGNs with $L_{\rm bol}$ of $10^{46-48}$ erg s$^{-1}$ at $z=1-2$.
On the contrary, the \citet{mcgill08} sample covers broad-line 
AGNs with $L_{\rm bol}$ of around $10^{45}$ erg s$^{-1}$ at $z\sim0.3$.
The distribution of the former samples are consistent each other and
follow similar trend of the SXDS luminous broad-line AGNs.
The latter sample has systematic offset from the former samples and
shows smaller H$\alpha$ broad-line FWHM than \ion{Mg}{2} FWHM.
The trend is similar to that seen in the SXDS less-luminous broad-line
AGNs.

We determine the relationship between the \ion{Mg}{2} and H$\alpha$ FWHMs
applying a BCES (Bivariate Correlated Error and intrinsic Scatter) 
bisector regression analysis \citep{akritas96} to
the 48 broad-line AGNs including both luminous and less-luminous AGNs. 
Considering the size of the sample, we do not divide the sample by
luminosity in the analysis. The resulting relationship is
\begin{eqnarray}
&\log& {\rm FWHM}_{\rm MgII} \nonumber \\
&=& (0.795\pm0.075) \times \log {\rm FWHM}_{\rm H\alpha} + (0.771\pm0.273). \nonumber \\
\end{eqnarray}
The rms scatter of $\log {\rm FWHM}_{\rm MgII}$ determined with the above
relationship using the measured $\log {\rm FWHM}_{\rm MgII}$ is 0.11 dex with a resulting
uncertainty for $\log M_{\rm BH}$ of 0.22 dex. 
The less luminous broad-line AGNs show systematic offset
of 0.1 dex on average from the relation.

The relationship between the FWHMs of the H$\beta$ and \ion{Mg}{2} broad-lines 
from \citet{mclure02} ({\it short dashed line}), \citet{onken08}
({\it thin solid line}), \citet{wang09} ({\it dotted line}), 
and \citet{croom11} ({\it long dashed line}) 
are also shown in the figure. These relationships for the
H$\beta$ FWHMs are converted to those for the H$\alpha$ FWHMs using the relationship
between the FWHMs of the broad H$\alpha$ and H$\beta$ lines determined
from $z<0.35$ for local broad-line AGNs of SDSS \citep[Eq.(3) of ][]{greene05}.
The relationship between the FWHMs of H$\alpha$ and H$\beta$ is consistent
with a recent determination of \citet{shen08b} but is slightly
offset from the relationship of \citet{schulze10}.
The distribution of SXDS broad-line AGNs
is consistent with the relationship of \citet{mclure02} and \citet{onken08}.
The relationship of \citet{croom11} has a steeper slope than the other
relationships and follows the distribution of SXDS broad-line AGNs except for the 
objects with the largest FWHMs. The relationship of \citet{wang09} 
shows a systematic offset from the other relationships and the distribution
of SXDS broad-line AGNs. The origin of the shift is unclear, 
but it is possible that \citet{wang09} use T06 \ion{Fe}{2} template,
and FWHM of \ion{Mg}{2} is measured systematically smaller than 
other measurements with VW01 \ion{Fe}{2} template (see Section~\ref{sec_Mg2}).

We use the relationship shown as the Equation (2)
to convert the H$\alpha$ FWHM to \ion{Mg}{2} FWHM and
the same black hole mass equation for \ion{Mg}{2} FWHM is applied. 
In total, we use the H$\alpha$ FWHM
measurements for 23 out of 116 broad-line AGNs at redshifts between 1.18 and 1.68.

\subsection{3000{\AA} Monochromatic Luminosities}\label{sec_UVOpt}

\begin{figure}
 \begin{center}
  \includegraphics[scale=0.85]{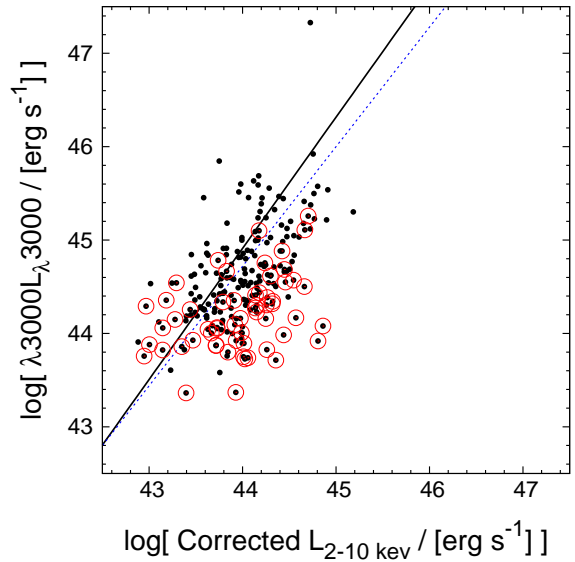}
  \caption{
Absorption corrected 2--10~keV luminosity vs. 3000{\AA} 
monochromatic luminosity of the SXDS broad-line AGNs ({\it filled circles}).
Broad-line AGNs redder than $R-i>0.3$ magnitude are marked 
with large {\it open circles}. The {\it thick solid line} indicates the relationship
determined by Marconi et al. (2004). The {\it dotted line} shows the
relationship determined with the BCES bisector analysis for broad-line AGNs
bluer than $R-i<0.3$ magnitude.
\label{L3000vsHX}}
 \end{center}
\end{figure}

For most of the objects, we derive 3000{\AA} monochromatic luminosity 
using the best-fit power-law continuum component described in 
Section~\ref{sec_Mg2}. We do not include the contribution of the
Balmer continuum in the fitting, the 3000{\AA} monochromatic luminosity
can be overestimated by 0.12 dex \citep{shen12b}.
Optical spectra covering rest-frame 3000{\AA} are not available for
broad-line AGNs that only covered by the NIR spectroscopic data.
We estimate their 3000{\AA} monochromatic luminosity by interpolating 
multi-band photometry data. All of the broad-line AGNs are 
detected in the deep multi-band images obtained with Suprime-Cam. 
We derive their rest-frame 3000{\AA} flux by interpolating the photometric 
measurements in the neighboring two bands around rest-frame 3000{\AA}.
The photometric data can include broad-emission lines and
Balmer continuum as well, and 
the 3000{\AA} luminosity may thus be affected by the broad-line component.
In order to estimate this effect, we compare the 3000{\AA} 
luminosity derived from normalized
spectra and multi-band photometry for objects with both measurements.
They are consistent each other within the rms of 0.10 dex and resulting
uncertainty of $M_{\rm BH}$ is 0.05 dex. It needs to be noted that
the optical spectra are normalized to match the $R$-band photometry from
the imaging observations as described in Section~\ref{sec_Mg2}, therefore the 
scatter only reflects the object-to-object variation of the strength
of the broad-line components. For the estimation of
the 3000{\AA} monochromatic luminosity with the photometric data, the
contribution from the Balmer continuum is not considered.
Uncertainty associated with the variability
of broad-line AGNs is already described in Section~\ref{sec_Mg2}.

For mildly obscured broad-line AGNs,
the 3000{\AA} luminosity can be affected by dust extinction.
Additionally for low-luminosity broad-line AGNs, the 3000{\AA} luminosity can be affected
by a host galaxy component. 
In such cases, the neither the 3000{\AA} luminosity derived from the spectra nor 
the photometry is a good indicator of the intrinsic UV luminosity.
In Figure~\ref{L3000vsHX}, the monochromatic 3000{\AA} luminosity
and absorption corrected 2--10~keV luminosity of the broad-line 
AGNs are shown. In this figure, broad-line AGNs with $R-i$ color
redder than 0.3 are marked with open circles. Those with a $R-i$ color of 0.3
are redder than the scatter of typical broad-line AGNs in the
redshift range observed in SDSS \citep{richards03}. Hereafter, we 
designate broad-line AGNs with $R-i$ redder (bluer) than 0.3 as red (blue)
broad-line AGNs. In the diagram, blue broad-line AGNs follow the relationship 
expected from the typical SEDs of broad-line AGNs as a function of intrinsic luminosity
which is shown with {\it solid line} in the figure \citep{marconi04}.
The {\it dotted line} in the figure shows the relationship determined with
the BCES bisector analysis for the blue broad-line AGNs.
In \citet{marconi04}, the SEDs around 3000{\AA} are described with a power-law with 
$\alpha=-0.44$ with $L_{\nu} \propto \nu^{\alpha}$ and a dependence on
optical-to-X-ray flux ratio, $\alpha_{\rm OX}$, on optical luminosity of broad-line 
AGNs \citep{vignali03} is considered. The $B$-band luminosity used in \citet{marconi04}
is converted to a 3000{\AA} monochromatic luminosity  assuming a
typical SED of broad-line QSOs 
\citep[$1.549 \times \lambda 4360 L_{\lambda 4360} = \lambda 3000 L_{\lambda 3000}$;][]{richards06}.
The consistency of the distribution with this relation suggests that the blue broad-line
AGNs have SEDs consistent with the optical-to-X-ray 
luminosity ratio, $\alpha_{\rm ox}$, dependence on luminosity for typical 
non-absorbed broad-line AGNs.

On the contrary, most of the
red broad-line AGNs have a systematically fainter 3000{\AA} monochromatic luminosity than
the blue broad-line AGNs at the same absorption corrected 2--10~keV luminosity.
The fainter 3000{\AA} luminosity suggests that the red broad-line AGN are affected by
mild dust absorption although most of them show a strong \ion{Mg}{2}
broad-line. Additionally, some of the red broad-line AGNs are brighter
in their 3000{\AA} luminosity. Most of them have the lowest absorption-corrected 2--10~keV luminosity
and their red colors can be explained by contamination by a host galaxy component 
in the wavelength range.
In both cases, the 3000{\AA} luminosity is not a good indicator of intrinsic
luminosity, thus we use absorption-corrected
X-ray luminosity instead of the 3000{\AA} monochromatic
luminosity for the black hole mass estimation. 
We convert the absorption-corrected hard X-ray luminosity
of the red broad-line AGNs to the intrinsic 3000{\AA} monochromatic
luminosity using the relationship derived by \citet{marconi04}. Considering the
scatter of blue broad-line AGNs around the relationship, we estimate the 
rms uncertainty of the intrinsic 3000{\AA} monochromatic luminosity
is 0.54 dex, which corresponds to 0.27 dex in the $M_{\rm BH}$ uncertainty. 
We use the 3000{\AA} monochromatic
luminosity derived from the hard X-ray luminosity of 59 red broad-line 
AGNs out of the 215 broad-line AGNs for the $M_{\rm BH}$ estimate. There are 26
red broad-line AGNs among the 116 broad-line AGNs in the redshift range 
between $1.18$ and $1.68$ where the broad-line AGN
BHMF and ERDF are derived below. 

In summary, 3000{\AA} monochromatic luminosities are derived
in three ways, from the power-law component of the fitting of
optical spectrum, the optical broad-band
photometry for broad-line AGNs only with the NIR spectrum, 
and hard X-ray luminosity for mildly-obscured or less-luminous
broad-line AGNs. 
The resulting 3000{\AA} monochromatic luminosities
are summarized in Table~\ref{tbl_FitRes}.

\section{Black Hole Mass, Bolometric Luminosity and Eddington Ratio}\label{sec_LE}

\begin{figure*}
 \begin{center}
 \includegraphics[scale=1.2]{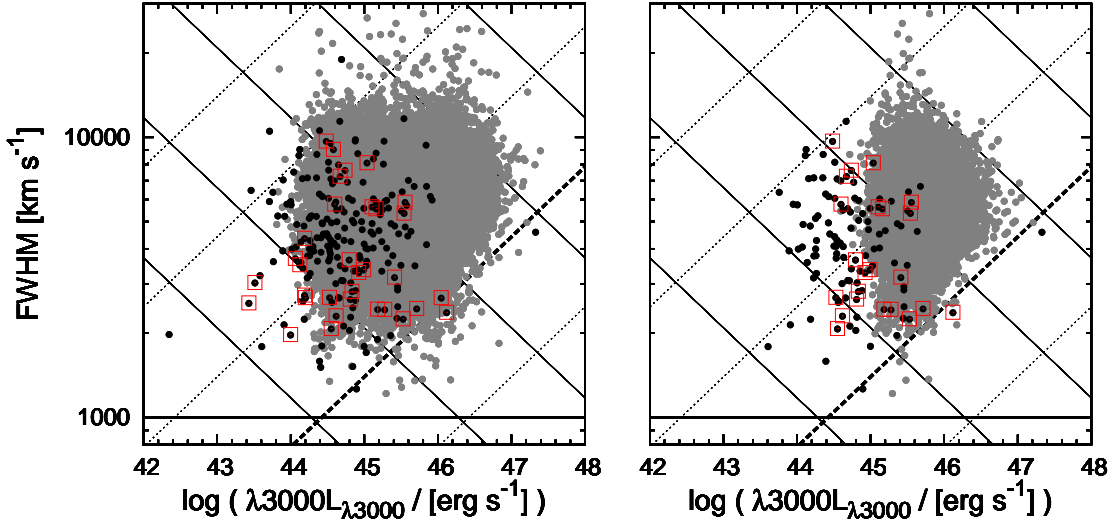}	
  \caption{
\ion{Mg}{2} FWHM vs. 3000{\AA} monochromatic luminosities of SXDS broad-line
AGNs ({\it filled circles}). Broad-line AGNs whose \ion{Mg}{2} FWHMs are estimated based
on H$\alpha$ FWHM are marked with {\it large open squares}.
The {\it solid lines} indicate constant $M_{\rm BH}$ values for
$M_{\rm BH}$ of $10^7$, $10^8$, $10^9$, $10^{10}$, and $10^{11}$ $M_{\sun}$
from bottom to top. The {\it dotted lines} show constant 
$\lambda_{\rm Edd}$ for $\log \lambda_{\rm Edd}$ of  $1$, $0$, $-1$, $-2$, and $-3$
from bottom to top. The {\it thick dotted line} indicates $\log \lambda_{\rm Edd}$ of $0$.
The {\it thick solid line} shows \ion{Mg}{2} FWHM of 1000 km s$^{-1}$.
SDSS DR5 broad-line AGNs that have a \ion{Mg}{2} FWHM measurement \citep{shen08a} are plotted as {\it gray circles}.
Right) Same for broad-line AGNs at $z=1.18-1.68$ only.
\label{fig_FWHM-3000}}
 \end{center}
\end{figure*}

\begin{figure}
 \begin{center}
  \includegraphics[scale=0.8]{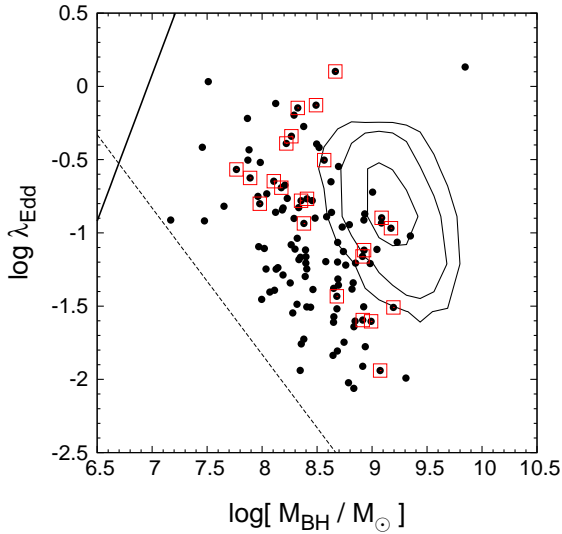}
  \caption{Black hole mass versus Eddington ratio of broad-line AGNs at redshifts
between 1.18 and 1.68.
Broad-line AGNs whose \ion{Mg}{2} FWHM are estimated from H$\alpha$ FWHM are marked with {\it large open squares}.
The {\it dotted line} shows the relationship between $\log M_{\rm BH}$ and
$\lambda_{\rm Edd}$ for a broad-line AGN with $L_{\rm 2-10~keV}=10^{43}$ erg s$^{-1}$, corresponding
to the faintest object in the sample.
The {\it thick solid line} indicate the constant \ion{Mg}{2} FWHM of 1000 km s$^{-1}$.
The distribution of the SDSS DR5 sample \citep{shen08a} in the 
same redshift range is shown with the contours.
\label{fig_M-E}}
 \end{center}
\end{figure}

The distribution of broad-line AGNs in the measured \ion{Mg}{2} FWHM vs 
intrinsic 3000{\AA} monochromatic luminosity plane is shown in 
the left panel of Figure~\ref{fig_FWHM-3000}. Large open squares indicate the 
FWHMs that are estimated with the broad H$\alpha$ line and Equation (2).
In the panel, the constant $M_{\rm BH}$ line derived from 
Equation (1) is shown with solid lines. From bottom to top, the
lines correspond to $M_{\rm BH}$ of $10^6$, $10^7$, $10^8$, $10^9$, $10^{10}$, 
and $10^{11}$ $M_{\sun}$. The SXDS broad-line AGNs 
cover the $M_{\rm BH}$ range between $10^{7} \sim 10^{10} M_{\sun}$ with a
median of $3.2 \times 10^8 M_{\sun}$.
 
Although we define broad-line AGNs as having FWHMs above 1000 km s$^{-1}$, 
a few objects have FWHMs between 1000 and 2000 km s$^{-1}$.
In the panel, we compare the distribution with that of broad-line AGNs from
SDSS DR5 \citep[filled gray circles;][]{shen08a}. The 49,526 SDSS broad-line AGNs,
which are selected with broad-line component whose FWHM is larger than 1200 km s$^{-1}$, 
are distributed between $z=0.3$ and $2.4$. Though there are far larger number of 
broad-line AGNs in the SDSS sample than in the SXDS sample, again only a
negligible fraction of broad-line AGNs have FWHMs smaller than 2000 km s$^{-1}$.
A similar rapid decrease of broad-line AGNs with FWHM smaller than 2000 km s$^{-1}$
is also reported by \citet{hao05} and \citet{stern12}. They select broad-line AGNs
in the local Universe from SDSS galaxy as well as quasar samples
with broad H$\alpha$ line above 1000 km s$^{-1}$.
The distribution of H$\alpha$ FWHMs shows a rapid decrease in the FWHM range 
below a few 1000 km s$^{-1}$. The physical origin of the cut off in the
FWHM distribution is unknown. 

In the right panel of the figure, only SXDS and SDSS broad-line AGNs at 
redshifts between 1.18 and 1.68 are plotted. The luminosity limits of the 
surveys define the left-hand envelopes of the distributions. 
It can be seen that the SXDS broad-line AGNs cover a 3000{\AA} 
monochromatic luminosity down to $10^{43.5}$ erg s$^{-1}$ in the redshift range.
This luminosity limit is more than an order of magnitude fainter 
than SDSS sample in the same redshift range.

We estimate the bolometric luminosity, $L_{\rm bol}$ 
from the 3000{\AA} monochromatic luminosity and using a bolometric correction 
factor of 5.8 for the monochromatic luminosity from \citet{richards06},
following \citet{vestergaard09}. 
In Table~\ref{tbl_ObjInfo}, $M_{\rm BH}$, $L_{\rm bol}$, and
$\lambda_{\rm Edd}$ are tabulated.
The bolometric luminosities of the SXDS sample 
are distributed between $L_{\rm bol} = 10^{45} \sim 10^{46}$ erg s$^{-1}$ with a median of 
$2.9 \times 10^{45}$ erg s$^{-1}$. The Eddington ratio of each object
is calculated as $\lambda_{\rm Edd} \equiv L_{\rm bol} / L_{\rm Edd}$. 
$L_{\rm Edd}$ is the Eddington-limited luminosity given by 
$1.26 \times 10^{38} \ M_{\rm BH}$ (erg s$^{-1}$) with $M_{\rm BH}$ in units of $M_{\sun}$.
The {\it dotted lines} in Figure~\ref{fig_FWHM-3000} represent constant $\log \lambda_{\rm Edd}$
values with $1$, $0$, $-1$, $-2$, $-3$, and $-4$ from bottom to top. 
$\log \lambda_{\rm Edd}$ of $0$ is shown with {\it thick dotted line}. The SXDS broad-line
AGNs range in $\log \lambda_{\rm Edd}$ between $0$ and $-2$. 

The uncertainties of $M_{\rm BH}$ for broad-line AGNs
with FWHM of \ion{Mg}{2} and 3000{\AA} monochromatic luminosity
of the spectroscopic data are typically 0.1 dex
from the FWHM measurement and 0.04 dex from 
the variability between the epochs of 
spectroscopic and imaging observations.
As we described in Section~\ref{sec_Mg2}, there can be systematic
uncertainty of 0.08 dex due to the \ion{Fe}{2} template uncertainty.
For broad-line AGNs with only $H\alpha$ measurement, 
the scatter between the FWHMs of \ion{Mg}{2} and H$\alpha$
and the uncertainty due to the 3000{\AA} monochromatic luminosity
add $M_{\rm BH}$ uncertainties of 0.22 and 0.05 dex, respectively.
Furthermore, for mildly-obscured or less-luminous 
broad-line AGNs whose 3000{\AA} monochromatic luminosity
is estimated with hard X-ray luminosity,
the additional $M_{\rm BH}$ uncertainty of 0.27 dex comes from the luminosity
conversion. Because $L_{\rm bol}$ is determined with $L_{\lambda 3000}$, which
is used for the $M_{\rm BH}$ estimation, the uncertainty of the $\lambda_{\rm Edd}$ 
is the same for the uncertainty of the $M_{\rm BH}$.

The distribution of broad-line AGNs at redshifts between 
1.18 and 1.68 in the $\lambda_{\rm Edd}$ and $M_{\rm BH}$ plane is shown in 
Figure~\ref{fig_M-E}.
Because of the flux limit of the survey, no object in the lower left
corner on the plane can be detected. It should be noted that the flux
limit runs diagonally. The {\it dotted line} in the figure shows the constant
hard X-ray luminosity line with $\log L_{\rm 2-10keV} ({\rm erg \ s}^{-1})=43$.
For $10^{8} M_{\sun}$ and $10^{7} M_{\sun}$,
we can detect objects with $\log \lambda_{\rm Edd}$ down to $-1.5$ and $-0.5$, 
respectively. The distribution is compared with those from SDSS ({\it contour}). 
Thanks to the deep detection limit of SXDS, we can select objects with 
lower mass as well as lower Eddington ratio in the same redshift range.
In the figure, the relation between $\lambda_{\rm Edd}$ and $M_{\rm BH}$
with constant FWHM of 1000 km s$^{-1}$ is shown with the {\it thick solid line}.
The cutoff in the distribution of FWHM seen in Figure~\ref{fig_FWHM-3000} appears as a deficit of
objects in the upper left region with $M_{\rm BH} < 10^{8} (M_{\sun})$ and 
$\log \lambda_{\rm Edd} > -0.5$, although the region is above 
the detection limit. 

\begin{deluxetable}{ccccc}
\tabletypesize{\footnotesize}
\tablecaption{Black hole mass and $\lambda_{\rm Edd}$ of broad-line AGNs \tablenotemark{a}\label{tbl_ObjInfo}}
\tablewidth{0pt}
\tablehead{
\colhead{ID} & \colhead{z} & \colhead{$\log[ M_{\rm{BH}} / M_{\odot}$]} & \colhead{ $\log[ L_{\rm{Bol}}$}  & \colhead{ $\log \lambda_{\rm Edd}$} \\ \colhead{ } & \colhead{ } & \colhead{ } & \colhead{/ [erg $\rm{s}^{-1}$] ]} & \colhead{ }
}
\startdata
0010     & 1.225 &  8.8 & 45.65 & $-$1.22 \\
0018     & 1.452 &  8.9 & 45.92 & $-$1.12 \\
0019     & 1.447 &  8.9 & 46.18 & $-$0.87 \\
0023     & 1.534 &  8.9 & 45.87 & $-$1.16 \\
0027     & 2.067 &  9.0 & 46.61 & $-$0.52 \\
0034     & 0.952 &  8.5 & 45.49 & $-$1.11 \\
0036     & 0.884 &  8.5 & 46.00 & $-$0.64 \\
0037     & 1.202 &  8.4 & 45.21 & $-$1.30 \\
0050     & 1.411 &  7.9 & 45.56 & $-$0.43 \\
0056     & 1.260 &  9.0 & 45.50 & $-$1.60 \\
\enddata
\tablenotetext{a}{Table 4 is published in its entirety in the electronic
edition of ApJ. A portion is shown here for guidance regarding its form
and content.}
\end{deluxetable}

\section{Black Hole Mass \& Eddington Ratio Distribution Functions}\label{sec_FUNC}

\subsection{{\it Binned} broad-line AGN BHMF and ERDF with $V_{\rm max}$ Method}\label{sec_NCMF}

\begin{figure}
 \begin{center}
  \includegraphics[scale=0.65]{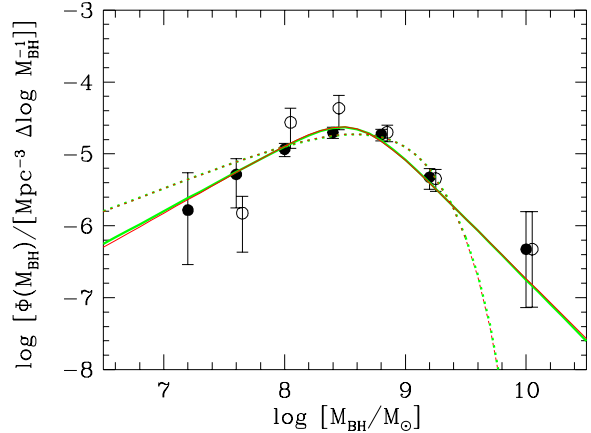}
  \caption{{\it Binned} active BHMFs for broad-line AGNs derived with
soft- ({\it filled circles}) and hard- ({\it open circles}) band samples.
{\it Lines} indicate {\it corrected} BHMF determined using the Maximum likelihood method
with four different combinations of functional forms of BHMF and ERDF.
{\it Solid lines} are for the double-power-law BHMF and
{\it dotted lines} are for the Schechter BHMF.
{\it Thick lines} are determined with the log-normal ERDF and
{\it thin lines} are determined with the Schechter ERDF.
Because the difference of the {\it corrected} BHMF with the different ERDFs 
is small, {\it thick} and {\it thin} lines overlap.
  \label{mass_plot0}}
 \end{center}
\end{figure}

\begin{figure}
 \begin{center}
  \includegraphics[scale=0.65]{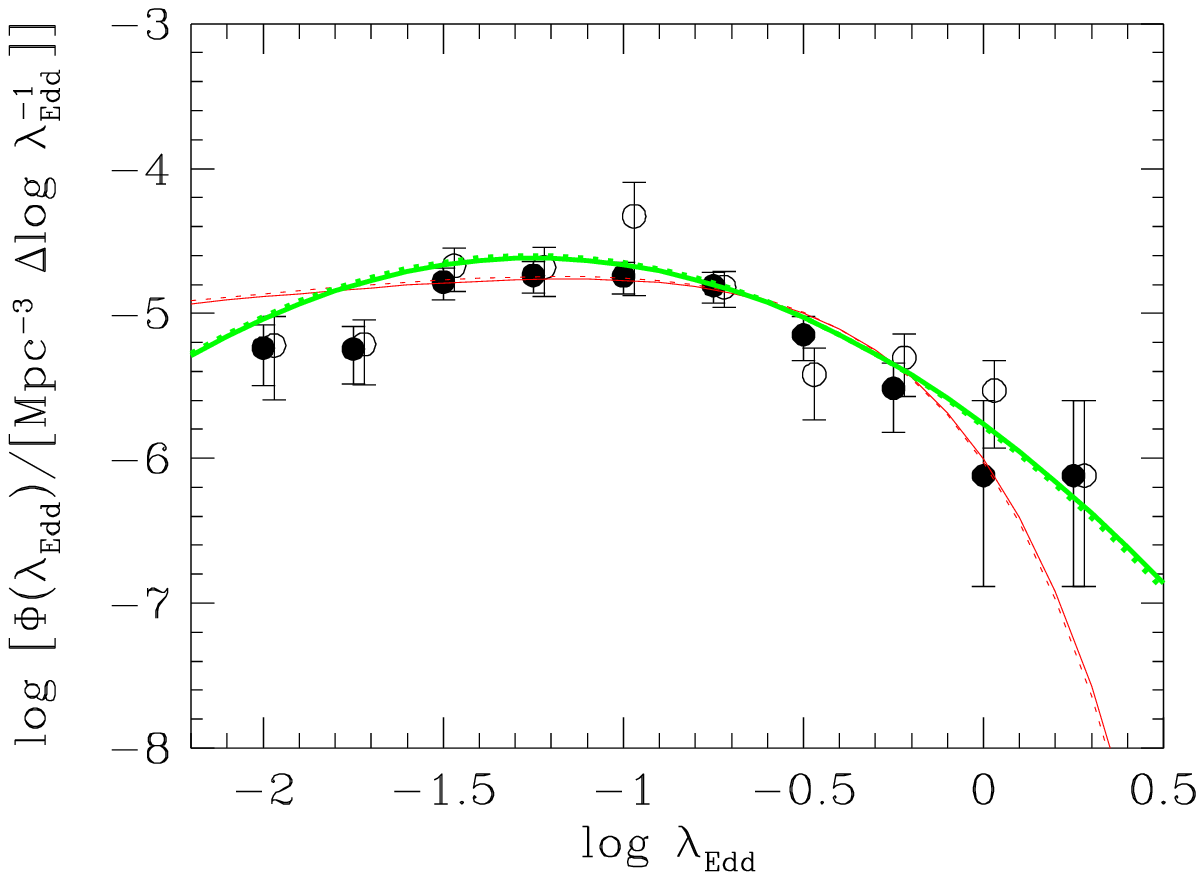}
  \caption{{\it Binned} ERDFs for broad-line AGNs derived with
soft- ({\it filled circles}) and hard- ({\it open circles}) band samples.
Lines indicate {\it corrected} ERDF determined using the Maximum likelihood method
with four different combinations of functional forms of BHMF and ERDF.
{\it Solid lines} are for the double-power-law BHMF and
{\it dotted lines} are for the Schechter BHMF.
{\it Thick lines} are for the log-normal ERDF and
{\it thin lines} are for the Schechter ERDF.
Because the difference of the {\it corrected} ERDFs with the different BHMFs 
is small, {\it solid} and {\it dotted lines} overlap.
All of the ERDFs are
normalized by matching the number density in the range
$\log \lambda_{\rm Edd}=-2.0 - 1.0$ to the number density of
broad-line AGN BHMF in the range $\log M_{\rm BH}=7.0 - 11.0$.
  \label{ER_plot0}}
  \end{center}
\end{figure}

\begin{figure}
 \begin{center}
  \includegraphics[scale=0.65]{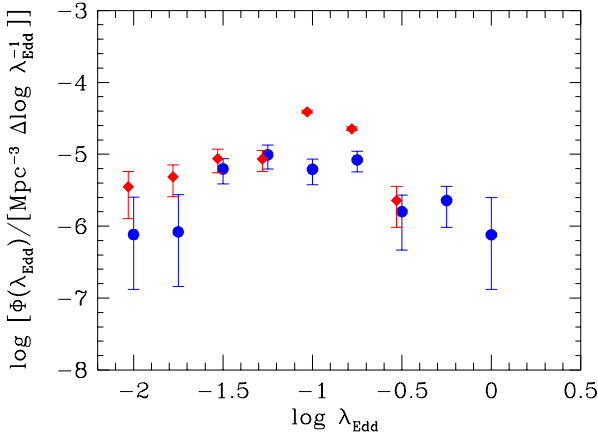}
  \caption{{\it Binned} ERDFs for soft-band broad-line AGNs
in the two mass ranges $10^{8.0-8.5}$ and $10^{8.5-9.0} M_{\sun}$
are shown with {\it filled circles} and {\it filled diamonds},
respectively. 
  \label{ER_plot0b}}
  \end{center}
\end{figure}

\begin{deluxetable*}{rrcrrrcrr}
\tabletypesize{\footnotesize}
\tablecaption{{\it Binned} broad-line AGN BHMF \label{tbl_MF_broad}}
\tablewidth{0pt}
\tablehead{
\colhead{$\langle \log [ M_{\rm BH} / M_{\sun} ]\rangle$ \tablenotemark{a}} &
\multicolumn{3}{c}{$\Phi_{\rm soft} \left( \log M_{\rm BH} \right)$ \tablenotemark{b}} &
\colhead{$N_{\rm soft}$} &
\multicolumn{3}{c}{$\Phi_{\rm{hard}} \left( \log M_{\rm BH} \right)$ \tablenotemark{b}} &
\colhead{$N_{\rm hard}$} \\
}
\startdata
 7.2 &  1.65 & $^{+}_{-}$ & $^{3.80}_{1.36}$\tablenotemark{a} &  1 & 
                        \multicolumn{3}{c}{\nodata}           &  0 \\
 7.6 &  5.17 & $\pm$      & 3.41             &  4 & 
        1.50 & $\pm$      & 1.07             &  2 \\
 8.0 & 11.60 & $\pm$      & 2.43             & 23 &
       27.46 & $\pm$      & 15.50            & 17 \\
 8.4 & 19.90 & $\pm$      & 3.55             & 36 &
       43.21 & $\pm$      & 21.54            & 30 \\
 8.8 & 18.62 & $\pm$      & 3.36             & 34 &
       19.93 & $\pm$      & 5.21             & 28 \\
 9.2 &  4.74 & $\pm$      & 1.50             & 10 &
        4.53 & $\pm$      & 1.51             &  9 \\
 9.6 &  \multicolumn{3}{c}{\nodata}          &  0 &
        \multicolumn{3}{c}{\nodata}          &  0 \\
10.0 &  0.47 & $^{+}_{-}$ & $^{1.09}_{0.39}$\tablenotemark{c} &  1 &
        0.47 & $^{+}_{-}$ & $^{1.09}_{0.39}$\tablenotemark{c} &  1 \\
\enddata
\tablenotetext{a}{The central value of ${\rm{M}}_{\rm{BH}}$ in each bin. The bin size is 0.4 dex and extends $\pm$ 0.2dex from the central value.}
\tablenotetext{b}{In units of $10^{-6} \left[ {\rm{Mpc}}^{-3} \left( \Delta \log [ M_{\rm BH} / M_{\sun}] \right)^{-1} \right] $}
\tablenotetext{c}{The upper and lower limits are determined following Gehrels (1986).}
\end{deluxetable*}

\begin{deluxetable*}{rrcrrrcrrrcrrrcrr}
\tabletypesize{\footnotesize}
\tablecaption{{\it Binned} ERDF at $z=1.43$ \label{tbl_EF}}
\tablewidth{0pt}
\tablehead{
\colhead{$\langle \log{ \lambda_{\rm Edd} }\rangle$\tablenotemark{a}} &
\multicolumn{3}{c}{$\Phi_{\rm soft} \left( \log{\lambda_{\rm{Edd}} } \right)$\tablenotemark{b}} &
\colhead{$N_{\rm{soft}}$} & 
\multicolumn{3}{c}{$\Phi_{\rm hard} \left( \log{\lambda_{\rm{Edd}}} \right)$\tablenotemark{b}} &
\colhead{$N_{\rm{hard}}$} \\
}
\startdata
 -2.00 &  5.79 & $\pm$      & 2.60                              &  6 &  6.03 & $\pm$      &  3.49             &  4 \\
 -1.75 &  5.68 & $\pm$      & 2.43                              &  6 &  6.12 & $\pm$      &  2.89             &  5 \\
 -1.50 & 16.51 & $\pm$      & 4.11                              & 18 & 21.29 & $\pm$      &  7.06             & 14 \\
 -1.25 & 18.36 & $\pm$      & 4.53                              & 20 & 20.80 & $\pm$      &  7.77             & 15 \\
 -1.00 & 18.15 & $\pm$      & 4.47                              & 20 & 46.98 & $\pm$      & 33.71             & 16 \\
 -0.75 & 15.54 & $\pm$      & 3.77                              & 18 & 15.23 & $\pm$      &  4.16             & 15 \\
 -0.50 &  7.12 & $\pm$      & 2.38                              & 10 &  3.79 & $\pm$      &  1.94             &  5 \\
 -0.25 &  3.04 & $\pm$      & 1.52                              &  4 &  4.95 & $\pm$      &  2.28             &  5 \\
  0.00 &  0.76 & $^{+}_{-}$ & $^{1.75}_{0.63}$\tablenotemark{c} &  1 &  2.95 & $\pm$      &  1.76             &  3 \\
  0.25 &  0.76 & $^{+}_{-}$ & $^{1.74}_{0.63}$\tablenotemark{c} &  1 &  0.76 & $^{+}_{-}$ &  $^{1.74}_{0.63}$\tablenotemark{c} &  1  \\
\enddata
\tablenotetext{a}{The central value of $\lambda_{\rm Edd}$ in each bin.}
\tablenotetext{b}{In units of $10^{-6} \left[ {\rm{Mpc}}^{-3} \left( \Delta \log [ \lambda_{\rm Edd} ] \right)^{-1} \right] $}
\tablenotetext{c}{The upper and lower limits are determined following Gehrels (1986).}
\end{deluxetable*}

First we derive the {\it binned} BHMF and ERDF for the broad-line AGNs between 
$1.18 \le z \le 1.68$ using the $V_{\rm max}$ method \citep{avni80}.
Detailed numbers for the sample are shown in Table~\ref{tbl_MFsample}. 
In the calculations of the {\it binned} BHMF and ERDF, we only consider broad-line AGNs with
$\lambda_{\rm Edd}$ larger than 0.01, and remove 2 broad-line AGNs below the limit.
We also remove 2 broad-line AGNs (SXDS0613 and SXDS0738) with neither \ion{Mg}{2} 
nor H${\alpha}$ FWHM measurements in the redshift range. We derive the
{\it binned} broad-line AGN
BHMF and ERDF for the soft- and hard-band samples separately.

For the {\it binned} broad-line AGN BHMF, we divide the mass range 
$6.6 \leq \log (M_{\rm BH} / M_{\sun}) \leq 10.2$ into 9 bins with
bin width, $\Delta \log M_{\rm BH}$, of 0.4 dex. The number density in a bin between
$(\log M_{\rm BH} - \Delta \log M_{\rm BH} / 2) \sim (\log M_{\rm BH} + \Delta \log M_{\rm BH} / 2)$
is given by,
\begin{eqnarray}
  \Phi_{\rm BH}( M_{\rm BH} ) \Delta \log M_{\rm BH} = \sum^n_{i = 1} \frac{1}{V_{a, i}}.
\end{eqnarray}
The summation is done for the $n$ broad-line AGNs in the mass bin.
$i$ is the index for a broad-line AGN in the mass bin. $V_{a,i}$ is the effective
survey volume for the $i$-th broad-line AGN in the comoving coordinate and its inverse 
represents the contribution of the broad-line AGN to the comoving number density of the mass bin.
$V_{a,i}$ is given by,
\begin{eqnarray}
V_{a,i} = \int^{z_{\rm max}}_{z_{\rm min}} \Omega( L_{{\rm x} i}, \log N_{{\rm H} i}, z^{\prime} ) \left( \frac{1+z^{\prime}}{1+z_{\rm cen}} \right)^k \frac{{\rm d}V}{{\rm d}z^{\prime}} {\rm d}z^{\prime}.
\end{eqnarray}
$z_{\rm min}$ and $z_{\rm max}$ represent the redshift range for the {\it binned} 
broad-line AGN BHMF (1.18 and 1.68, respectively). 
$\Omega (L_{{\rm X} i}, \log N_{{\rm H} i}, z^{\prime})$ is the survey area which
is calculated assuming that the $i$-th broad-line AGN observed at $z_i$ with absorption-corrected
luminosity $L_{{\rm X} i}$, absorption hydrogen column density $\log N_{{\rm H} i}$ is
at redshift $z^{\prime}$. With an estimated absorption-corrected $L_{{\rm X} i}$ 
and $\log N_{{\rm H} i}$ as described in Section~\ref{sec_SAMPLE},
we calculate the predicted count-rate for each broad-line AGN with $z^{\prime}$
instead of the observed $z_i$. The same X-ray spectral model
of AGNs is used, as explained in Section~\ref{sec_SAMPLE}. The survey area for the sample with
likelihood larger than 7 is determined as a function of count-rate for the 
overlapping region of the X-ray and deep optical surveys \citep{ueda08}. 
The logN-logS relation derived with the area curve for the SXDS X-ray sources
is consistent with the relation determined in deeper {\it Chandra} surveys \citep{ueda08}.
The consistency implies the position-dependent detection limit
of the X-ray survey is reproduced well in the estimated area curve.
If the predicted count-rate of an object at a certain redshift $z^{\prime}$ 
is below the smallest count-rate limit of the survey, 
$\Omega (L_{{\rm X} i}, \log N_{{\rm H} i}, z^{\prime})$ becomes 0 above that redshift.
The factor $((1+z^{\prime})/(1+z_{\rm cen}))^k$  corrects for the number density
evolution with $(1+z)^{k}$ within the redshift range to determine the corresponding
number density at $z_{\rm cen}$ of 1.43. However, the  correction factor is negligible
even if we introduce rather strong number density evolution with $k=4-5$ observed
for the X-ray luminosity function of AGNs \citep{ueda03, hasinger05}, because the redshift
range for this calculation is narrow. 
Therefore, we neglect this term hereafter and fix the value $k$ as 0 
for simplicity. Finally $V_{a,i}$ is obtained by integrating the corresponding survey 
area for the object 
at redshift $z^{\prime}$ multiplied by the cosmological volume element 
$(dV / dz^{\prime}) dz^{\prime}$ of unit solid angle in the redshift range. 
The uncertainty of the {\it binned} broad-line AGN
BHMF is estimated using Poisson statistics as,
\begin{eqnarray}
 \sigma = \left[ \sum^n_{k=1} \left( \frac{1}{V_{a,i} \ \Delta \log M_{\rm BH}} \right)^2 \right]^{1/2}.
\end{eqnarray}
The resulting {\it binned} broad-line AGN
BHMFs are shown in Figure~\ref{mass_plot0} and
Table~\ref{tbl_MF_broad}. The filled and
open circles represent the {\it binned} BHMFs of soft- and hard-band 
broad-line AGN samples, respectively.
Both of the {\it binned} broad-line AGN BHMFs peak at around 
$M_{\rm BH}$ of $10^{8.5} M_{\sun}$.
The {\it binned} broad-line AGN
BHMFs are consistent each other within the 1 $\sigma$ uncertainty. 

The {\it binned} broad-line AGN
ERDF is derived in the same way for the {\it binned} BHMF 
using the $V_{\rm max}$ method.
We bin the sample in $\lambda_{\rm Edd}$ by dividing the $\lambda_{\rm Edd}$ range
of $-2.125 < \log \lambda_{\rm Edd} <  0.375$ into 10 bins with a bin
width $\Delta \lambda_{\rm Edd}$ of 0.25 dex. The number density of a certain
Eddington ratio bin between
( $\log \lambda_{\rm Edd} - \Delta \log \lambda_{\rm Edd} / 2 ) \sim ( \log \lambda_{\rm Edd} + \Delta \log \lambda_{\rm Edd} / 2 )$
is given by,
\begin{eqnarray}
 \Phi_{\lambda}( \lambda_{\rm Edd} ) \Delta \log \lambda_{\rm Edd} &=& \sum^n_{i = 1} \frac{1}{V_{a, i}}.
\end{eqnarray}
$n$ is the number of broad-line AGNs in the Eddington ratio bin. $V_{a, i}$ is the same
effective volume for the $i$-th broad-line AGN used in the binned BHMF. The uncertainty for
each Eddington ratio bin is again given by
\begin{eqnarray}
 \sigma &=& \left[ \sum^n_{k=1} \left( \frac{1}{V_{a,i} \Delta \log \lambda_{\rm Edd}} \right)^2 \right]^{1/2}.
\end{eqnarray}
The resulting {\it binned} ERDF is shown in Figure~\ref{ER_plot0} and 
tabulated in Table~\ref{tbl_EF}. 
The filled and open circles in the figure are the {\it binned} ERDFs for the soft- and hard- band 
samples, respectively. The {\it binned} ERDFs are consistent 
with each other within the 1 $\sigma$ uncertainty. 
In the {\it binned} ERDFs, broad-line AGNs in the entire mass range are considered.
Due to the flux limit of the survey, the sample covers only broad-line AGNs with larger $M_{\rm BH}$
in the lower $\lambda_{\rm Edd}$ bins. Therefore the shape of the {\it binned} ERDF can 
be affected by the flux limit. 

In order to examine the $M_{\rm BH}$ dependence of the {\it binned} ERDF, in Figure~\ref{ER_plot0b},
the {\it binned} ERDFs derived with the soft-band selected broad-line AGNs in 
the $M_{\rm BH}$ range between $10^{8.0-8.5} M_{\sun}$ 
and that between $10^{8.5-9.0} M_{\sun}$ are plotted with {\it filled circles}
and {\it filled diamonds}, respectively.
The overall shapes of the {\it binned} ERDFs do not significantly differ from each other within
the uncertainty, except
for the lowest $\log \lambda_{\rm Edd}$ range where the lower mass {\it binned} ERDF can be
affected by the flux limit. Due to the limited number and mass range of the SXDS sample,
there is no signature of the dependence of the ERDF on $M_{\rm BH}$.

\subsection{{\it Corrected} broad-line AGN
BHMF and ERDF with Maximum Likelihood Method}\label{sec_correction}

\begin{figure}
 \begin{center}
  \includegraphics[scale=0.65]{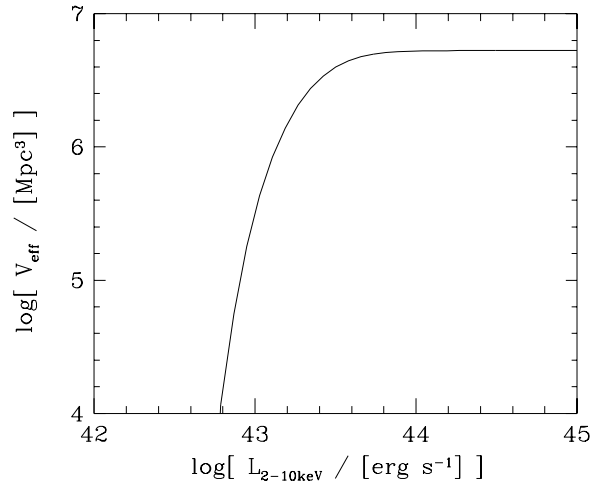}
  \caption{Effective volume of the SXDS soft-band sample in the redshift
range between 1.18 and 1.68 as a function of 2--10~keV hard X-ray luminosity.
  \label{Lx_vmax}}
  \end{center}
\end{figure}

\begin{deluxetable*}{llrrrrrrrl}
\tabletypesize{\footnotesize}
\tablecaption{Best fit parameters for the {\it corrected} broad-line AGN
BHMF and ERDF \label{tbl_parameter}}
\tablewidth{0pt}
\tablehead{
  \colhead{BHMF\tablenotemark{a}} &
  \colhead{ERDF\tablenotemark{b}} & 
  \colhead{$\phi^{*}$\tablenotemark{c}} & 
  \colhead{$\log M^{*}$\tablenotemark{d}} &
  \colhead{$\alpha$} & 
  \colhead{$\beta$} & 
  \colhead{$\log \lambda_{\rm Edd}^{*}$\tablenotemark{e}} & 
  \colhead{$\alpha_{\lambda}$\tablenotemark{f}} & 
  \colhead{2DKS\tablenotemark{g}} &
  \colhead{Note\tablenotemark{h}}
}
\startdata
DPL & SCH & $19.82_{-3.20}^{+3.20}$ & $8.57_{-0.15}^{+0.16}$ & $-0.05_{-0.28}^{+0.35}$ & $-2.67_{-0.39}^{+0.28}$ & $-0.58_{-0.10}^{+0.12}$ & $-0.74_{- 0.15}^{+ 0.15}$ & 86 & \\
SCH & SCH & $20.80_{-3.20}^{+3.30}$ & $8.77_{-0.08}^{+0.08}$ & $-0.35_{-0.14}^{+0.15}$ &  \nodata                & $-0.59_{-0.10}^{+0.11}$ & $-0.74_{- 0.15}^{+ 0.15}$ & 65 & $\times$ \\
DPL & LOG & $19.39_{-3.20}^{+3.20}$ & $8.59_{-0.15}^{+0.16}$ & $-0.09_{-0.27}^{+0.34}$ & $-2.70_{-0.40}^{+0.29}$ & $-2.87_{-0.20}^{+0.16}$ & $1.25_{-0.11}^{+0.14}$ & 79 & \\
SCH & LOG & $20.61_{-3.20}^{+3.30}$ & $8.78_{-0.08}^{+0.08}$ & $-0.35_{-0.14}^{+0.15}$ &  \nodata                & $-2.88_{-0.20}^{+0.16}$ & $1.24_{- 0.10}^{+0.13}$ & 62 & \\
\enddata
\tablenotetext{a}{DPL: double-power-law, SCH: Schechter functions.}
\tablenotetext{b}{SCH: Schechter, LOG: log-normal distributions.}
\tablenotetext{c}{In unit of $10^{-6}$ Mpc$^{-3}$ per unit $\log M_{\rm BH}$ bin.}
\tablenotetext{d}{In unit of $M_{\sun}$.}
\tablenotetext{e}{$\mu$ for log-normal distribution.}
\tablenotetext{f}{$\sigma$ for log-normal distribution.}
\tablenotetext{g}{2DKS probability in unit of \%.}
\tablenotetext{h}{Consistency with high-luminosity end of hard X-ray luminosity function. In details, see Section 5.3.}
\end{deluxetable*}

Both the {\it binned} broad-line AGN BHMF and ERDF are affected by 
the detection limit determined by the X-ray count rate; 
at the low-mass end of the {\it binned} BHMF the sample covers only high $\lambda_{\rm Edd}$ 
broad-line AGNs and at the low Eddington ratio end of the {\it binned} ERDF the sample
does not include broad-line AGNs with low $M_{\rm BH}$.  
Such detection limits are not corrected for in the calculations of the
{\it binned} broad-line AGN BHMF and ERDF. 

The effects of the detection limit can be corrected 
through statistical methods assuming the forms of  both functions
\citep{kelly09, kelly10, schulze10, shen12, kelly12}. \citet{schulze10} apply the
Maximum likelihood method to a sample of low-redshift broad-line AGNs 
detected in the ESO/Hamburg survey, assuming that the intrinsic 
ERDF does not depend 
on $M_{\rm BH}$. On the other hand, \citet{kelly10},
\citet{shen12} and \citet{kelly12} apply a Bayesian approach \citep{kelly09} to SDSS broad-line AGNs. 
They introduce a dependence of ERDF on the black hole mass.
Furthermore, the statistical scatter in the virial black hole mass estimate 
is considered in the calculations; the scatter can broaden a peak in the 
intrinsic broad-line AGN BHMF and a steep slope of 
intrinsic BHMF at the high-mass end 
can become flatter in the {\it binned} BHMF. 
If the intrinsic BHMF has a peak at a
certain black hole mass and there is a turn-over at
the low-mass end, the scatter can also affect the
{\it binned} BHMF in the low-mass end.

Here, we apply Maximum likelihood method used in \citet{schulze10} for the
broad-line AGNs in the soft-band sample.
Because there is no significant difference between the {\it binned} 
broad-line AGN ERDFs in the
high- and low-mass ranges as shown in Figure~\ref{ER_plot0b}, we assume that
the intrinsic broad-line AGN ERDF is constant regardless of black hole mass.
The effect of the scatter of the virial black hole mass estimate 
is not considered in this paper, because the SXDS sample covers smaller black hole
mass and Eddington ratio ranges than the SDSS sample, most of the
SXDS broad-line AGNs lie in the mass range below the knee of the BHMF,
and in order not to be affected by the uncertainty associated with the
modeling of the scatter.
The high-mass end of the {\it corrected} broad-line AGN
BHMF can be affected by the flattening due to the scatter.

In this evaluation, we assume the shape of the {\it corrected} 
broad-line AGN BHMF to be either
a double-power-law or a Schechter function,
\begin{eqnarray} 
&\Phi_{\rm BH}(M_{\rm BH})&\nonumber \\
&=& \frac{M_{\rm BH}}{\log_{10}e} \frac{\phi^{*}}{M^{*}} \left( \frac{1}{(M_{\rm BH}/M^{*})^{-\alpha}+(M_{\rm BH}/M^{*})^{-\beta}} \right), \nonumber \\ 
\\
&\Phi_{\rm BH}(M_{\rm BH})& \nonumber \\
&=& \frac{M_{\rm BH}}{\log_{10}e} \frac{\phi^{*}}{M^{*}} \left( \frac{M_{\rm BH}}{M^{*}} \right)^{\alpha} \exp \left( - \frac{M_{\rm BH}}{M^{*}} \right), 
\end{eqnarray}
respectively, with the functions expressed per $\log M_{\rm BH}$. We introduce
these two forms because the double-power-law describes the AGN 
luminosity function well, and the Schechter function describes the 
luminosity and mass functions of galaxies. Even though a modified Schechter 
function describes the non-active BHMF in the local Universe
\citep{aller02, shankar04}, we do not implement such a function,
since it does not converge and results
in a rather small $M^{*}$ ($10^{5-6} M_{\sun}$).

For the {\it corrected} ERDF we assume the Schechter function 
and the log-normal distributions.
\begin{eqnarray} 
&\Phi_{\lambda}(\lambda_{\rm Edd})& \nonumber \\
&=& \frac{\lambda_{\rm Edd}}{\log_{10}e} \frac{\phi^{*}_{\lambda}}{\lambda_{\rm Edd}^{*}} \left( \frac{\lambda_{\rm Edd}}{\lambda_{\rm Edd }^{*}} \right)^{\alpha_{\lambda}} \exp \left( - \frac{\lambda_{\rm Edd}}{\lambda_{\rm Edd}^{*}} \right), \\
&\Phi_{\lambda}(\lambda_{\rm Edd})& \nonumber \\
&=& \frac{1}{\lambda_{\rm Edd}\sqrt{2 \pi \sigma^{2}}} \exp \left( - \frac{({\rm ln}\lambda_{\rm Edd} - \mu)^{2}}{2 \sigma^{2}} \right). 
\end{eqnarray}
We do not consider the $M_{\rm BH}$ dependence of the ERDF. For both distributions, we 
define the normalized {\it corrected} ERDF with
\begin{eqnarray}
 P_{\lambda}(\lambda_{\rm Edd}) = \frac{\Phi(\lambda_{\rm Edd})}{ \displaystyle \int_{-2.0}^{1.0} \Phi(\lambda_{\rm Edd}) {\rm d} \log \lambda_{\rm Edd}}.
\end{eqnarray}
We normalize the {\it corrected} ERDF in the range $\log \lambda_{\rm Edd}$ of $-2.0$ and $1.0$.
In these functions, $\phi^{*}$, $\alpha$, $\beta$, and $M^{*}$ of the {\it corrected} BHMF and $\alpha_{\lambda}$ and $\lambda_{\rm Edd}^{*}$ of the {\it corrected} ERDF 
(or $\sigma$ and $\mu$ for log-normal ERDF) are free parameters. 
We derive the best fit parameters with the Maximum Likelihood method \citep{marshall83}.
The likelihood function is written as
\begin{eqnarray}
 S &=& - 2 \sum_{i=1}^{N} {\rm ln}\ p(M_{{\rm BH},i}, \lambda_{{\rm Edd},i}) \nonumber \\
   &+& 2\int\int p(M_{\rm BH}, \lambda_{{\rm Edd}})\ d\log \lambda_{\rm Edd}\ d \log M_{\rm BH}  
\end{eqnarray}
and the model parameters that minimize $S$ are the best fit parameters \citep{marshall83}. 
The sum of the first term will be taken for the entire $N$ broad-line AGNs in the sample.
The term $p(M_{\rm BH}, \lambda_{\rm Edd})$ is the expected number of black holes with 
$M_{\rm BH}$ and $\lambda_{\rm Edd}$ in unit $\log M_{\rm BH}$ and $\log \lambda_{\rm Edd}$ intervals 
in the survey redshift range
with the assumed $\Phi_{\rm BH}(M_{\rm BH})$ and $P_{\lambda}(\lambda_{\rm Edd})$. $p(M_{\rm BH}, \lambda_{\rm Edd})$ is
given by
\begin{eqnarray}
&p(M_{\rm BH},\lambda_{\rm Edd})& \nonumber \\
&=&\int_{z_{\rm min}}^{z_{\rm max}} \Omega^{\prime}(M_{\rm BH}, \lambda_{\rm Edd}, z^{\prime}) \nonumber \\
&& P_{\lambda}(\lambda_{\rm Edd}) \Phi_{\rm BH}(M_{\rm BH}) \left( \frac{1+z^{\prime}}{1+z_{\rm cen}} \right)^{k}\frac{{\rm d}V}{{\rm d}z^{\prime}}{\rm d}z^{\prime}. \nonumber \\
\end{eqnarray}
$\Omega^{\prime}(M_{\rm BH}, \lambda_{\rm Edd}, z^{\prime})$ is the survey area for
an broad-line AGN with $M_{\rm BH}$ and $\lambda_{\rm Edd}$ at $z^{\prime}$. We calculate
the expected count rate for each combination of $M_{\rm BH}$, $\lambda_{\rm Edd}$, and $z^{\prime}$
assuming the same model for the X-ray spectrum of AGNs used in Section~\ref{sec_SAMPLE}.
We convert the $L_{\rm bol}$ for a combination of $M_{\rm BH}$ and
$\lambda_{\rm Edd}$ to $L_{\rm 2-10~keV}$ with the bolometric correction factor
from \citet{marconi04}.
In this calculation we do not consider the effect of X-ray absorption because the 
effect is not significant for the sample of broad-line AGNs. 
In Figure~\ref{Lx_vmax}, we plot the effective volume
\begin{eqnarray}
V_{\rm eff}=\int_{z_{\rm min}}^{z_{\rm max}} \Omega^{\prime}(M_{\rm BH}, \lambda_{\rm Edd}, z^{\prime}) \frac{{\rm d}V}{{\rm d}z^{\prime}}{\rm d}z^{\prime} 
\end{eqnarray}
as a function of $L_{\rm 2-10~keV}$.
We consider $\log \lambda_{\rm Edd}$ down to $-2.0$ and $M_{\rm BH}$ of $10^{6} M_{\sun}$.
Again we use $k$ of 0 as explained in the previous subsection.
We minimize $S$ with the 6 free parameters with the downhill simplex
algorithm \citep{nelder65}.
The 1$\sigma$ uncertainty of the best-fit parameters for each model can be 
evaluated by the increase of $S$ by 1 from the minimum value. 
In order to determine the 1$\sigma$ uncertainty 
of each parameter, the parameter is changed from 
its best value to a different value, and fixing the parameter at the value, 
the same minimization process is applied 
for the other parameters and we evaluate the change of the minimum $S$ 
value from the best fit value of $S$. 
The uncertainty of the selected parameter is determined by the
change of the minimum of 1 from the best fit value.
The resulting best-fit parameters are summarized in Table~\ref{tbl_parameter}.

The resulting {\it corrected} broad-line AGN
BHMFs and ERDFs are shown in Figures~\ref{mass_plot0}
and \ref{ER_plot0}. The {\it corrected} ERDFs are normalized by matching the number 
density in the range $\log \lambda_{\rm Edd}=-2.0 - 1.0$ and that of
the {\it corrected} BHMF in the range $\log M_{\rm BH}=7.0 - 11.0$.
The {\it corrected} BHMF follows the estimated number density with the $V_{\rm max}$
method well in the mass range above $10^{8.5} M_{\sun}$.
This is consistent with the limit of the SXDS sample; it extends
down to a Eddington ratio of 0.01 in the mass range above $\sim10^{8} M_{\sun}$. 
In the lower mass range, the {\it corrected} BHMF is slightly larger than the
number density derived with the $V_{\rm max}$ method. The estimated 
correction is consistent with the {\it corrected} ERDF; for 
10$^{7.5} M_{\sun}$ black holes the sample covers a Eddington ratio of 
$\sim0.1$, and the  ratio below and above this Eddington 
ratio is $1.7-1.8$ from the {\it corrected} ERDF. Therefore, the estimated 
correction is not large. The shape of the {\it corrected} ERDF is
consistent with that derived in the $10^{8}-10^{9} M_{\sun}$
mass range with the $V_{\rm max}$ method. The correction required
for the ERDF is only significant in the low Eddington 
range, $\log \lambda_{\rm Edd}$ below $-1.5$. Thanks to the coverage for 
relatively low-luminosity broad-line AGNs, the SXDS sample
is rather complete in a wide mass and Eddington ratio range.

The minimum value of the likelihood function does not reflect 
the goodness of the fit. We evaluate the goodness of the fit by applying 
the two-dimensional Kolmogorov-Smirnov (2DKS) test on the
distribution of the sample in the 
$M_{\rm BH}$ and $\lambda_{\rm Edd}$ plane \citep{fasano87}.
The resulting 2DKS probability for the deviation in the plane
are shown in Table~\ref{tbl_parameter}. All of them 
exceed 20\% and all of the models fit well the distribution of
objects on the plane.

In the mass range below $10^{8.5}$M$_{\sun}$, the
{\it corrected} BHMFs show possible decline to the low-mass end.
The mass range is well above the detection limit as shown in 
Figure~\ref{fig_M-E}, but the decline can be affected by the
completeness of the broad-line AGN sample. For example, there is a 
possibility that among objects without spectroscopic identification
low-luminosity broad-line AGNs whose SEDs are
dominated by host galaxy components are classified as narrow-line
AGNs in the photometric redshift estimation, and they would be
missed in the current broad-line AGN sample.

\subsection{Constraint from the Hard X-ray Luminosity Function}\label{sec_luminosityfunc}

\begin{figure*}
 \begin{center}
  \includegraphics[scale=0.9]{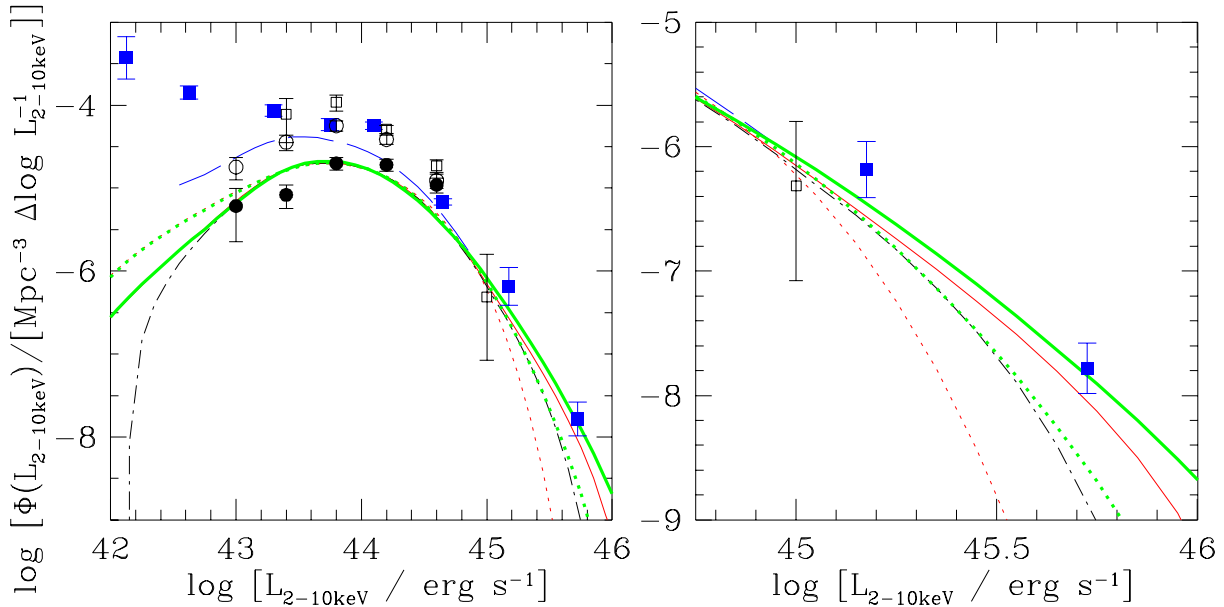}
  \caption{Left) Hard X-ray luminosity function of 
X-ray-selected AGNs (Ueda et al. 2012, in preparation; {\it filled squares}) compared
with that reproduced from the {\it corrected} broad-line AGN
BHMFs and ERDFs
({\it lines}). The {\it open circles} and {\it squares} show the luminosity
function derived from the soft-band-selected and hard-band-selected AGN samples of SXDS in the range
$1.18<z<1.68$, respectively. {\it Filled circles} represent the luminosity
function of broad-line AGNs in the soft-band sample, which
is used to derive the {\it corrected} broad-line AGN BHMF and ERDF.
The hard X-ray luminosities of all AGNs are corrected for
intrinsic absorption. Solid and dotted lines 
correspond to the luminosity functions for the BHMF characterized by  
double-power-law and Schechter functions 
and the thick and thin lines are those derived with the Schechter 
and log-normal ERDF, respectively. The {\it thin
dot-dashed line} shows the luminosity function derived with
double-power-law BHMF $+$ log-normal ERDF with limiting
$M_{\rm BH}$ range below $10^{10} M_{\sun}$.
The {\it thin long-dashed line} represents the luminosity function
with double-power-law BHMF $+$ log-normal ERDF after correcting for the
obscured fraction following Hasinger (2008).
Right) Zoom in version of the left panel in the high-luminosity range. 
  \label{expectLF}}
 \end{center}
\end{figure*}

The luminosity function of broad-line AGNs is the convolution of 
their BHMF and ERDF, 
therefore we can constrain the shapes of broad-line AGN BHMF and ERDF further by
using the luminosity function determined from a combination of various
AGN samples. In particular, the number density at the bright end of the
luminosity function obtained through wider but shallower surveys than
SXDS can constrain the shapes of broad-line AGN BHMF and ERDF in the high 
$M_{\rm BH}$ and $\lambda_{\rm Edd}$ range. It needs to be noted that the 
fraction of obscured narrow-line AGNs is low in the luminosity 
range \citep{hasinger08} and the luminous end of the luminosity function 
is thought to be dominated by broad-line AGNs. 

In Figure~\ref{expectLF}, we plot the observed hard X-ray luminosity 
function of AGNs at $z=1.2-1.6$ based on a combined sample of
AGNs from various hard X-ray surveys as {\it filled squares} 
(Ueda et al. 2012, in preparation). The horizontal
axis is the absorption-corrected 2--10~keV luminosity. The hard
X-ray luminosity functions of the broad-line + narrow-line AGNs 
in the SXDS soft-band-selected and hard-band-selected
samples are shown as {\it open circles} and {\it squares}, respectively. 
The hard-band-selected AGN luminosity function in 
the SXDS is consistent with that derived from the combined sample.
The soft-band-selected AGN luminosity function has slightly lower
number density than the hard-band-selected AGN luminosity function
below $L_{\rm 2-10~keV}=10^{44}$ erg s$^{-1}$. The
lower number density can be qualitatively explained with the fact that the
soft-band-selected sample misses heavily obscured AGNs and the fraction
of obscured AGNs is larger for AGNs with lower luminosity. Because
all of the luminosity functions are consistent each other in the 
luminosity range above $L_{\rm 2-10~keV}=10^{44}$ erg s$^{-1}$,
therefore we use the number density of AGNs at the luminous end of the
combined sample as the constraint on the broad-line AGN BHMF and ERDF.

The {\it filled circles} represent the hard X-ray luminosity function 
of broad-line AGNs in the soft-band samples. Again the hard X-ray luminosity
is corrected for intrinsic absorption. The hard X-ray luminosity 
function shows turn over at $L_{\rm 2-10~keV}=10^{44}$ (erg s$^{-1}$). Such
turn over is also observed in the broad-line AGN hard X-ray luminosity function
at $z=1-3$ from {\it Chandra} surveys \citep{yencho09}. At least
part of the decrease can be explained with the increasing fraction 
of obscured narrow-line AGNs in the 
lower luminosity range as described below. The lines in the figure show the 
hard X-ray luminosity function derived from the convolution of the 
{\it corrected} broad-line BHMF and 
ERDF. {\it Solid} and {\it dotted lines} correspond
to luminosity functions with the {\it corrected} BHMF of double-power-law and Schechter forms
and the {\it thick} and {\it thin lines} are those 
derived with Schechter and log-normal ERDF, respectively. 
The hard X-ray luminosity for each set of $M_{\rm BH}$ and $\lambda_{\rm Edd}$ 
is derived with the bolometric correction for 3000{\AA} monochromatic 
luminosity and the relation between $\lambda 3000 L_{\lambda 3000}$ 
and $L_{\rm 2-10keV}$ shown with the solid line in Figure~\ref{L3000vsHX}.
Considering the upper limit of the observed $M_{\rm BH}$ of broad-line
AGNs \citep{vestergaard08} and local galaxies \citep{mcconnell11}, 
we limit the $M_{\rm BH}$ range to be $10^{6-10.5} M_{\sun}$ and the
$\log \lambda_{\rm Edd}$ range to be $-2$ and $1$.

The hard X-ray luminosity functions derived with the double-power-law BHMF
or log-normal ERDF can reproduce the observed number density at the
high-luminosity end. On the contrary, if both the BHMF and ERDF are
modeled with an exponential-cutoff such as the Schechter BHMF with Schechter ERDF
({\it thin dotted line}),
the high-luminosity end of the luminosity function cannot be reproduced;
the predicted number density is more than one order of magnitude smaller
than the observed luminosity function. Therefore such models are unlikely 
to represent the BHMF and ERDF of broad-line AGNs at $z=1.4$.  
Furthermore, if we limit the mass range of BHMF up to $10^{10} M_{\sun}$, 
the predicted luminosity function with a double-power-law BHMF and log-normal ERDF
are represented by the dot-dashed line in the figure; the predicted density 
in the high luminosity end is much lower than 
the observed one. Thus in order to reproduce the number density of luminous AGNs,
the BHMF needs to be extended up to $10^{10.5} M_{\sun}$ under the assumption
that the ERDF is constant over the wide $M_{\rm BH}$ range.

In the luminosity range below $L_{\rm 2-10~keV}=10^{44}$ erg s$^{-1}$,
the broad-line AGN luminosity function shows decline toward lower luminosity. 
At least part of the decline can be explained with the increasing fraction 
of obscured narrow-line AGNs in the lower luminosity range.  In 
Figure~\ref{expectLF}, we also plot model luminosity function derived
with a double-power-law BHMF and log-normal ERDF after correcting 
the fraction of narrow-line AGN following \citet{hasinger08} with
{\it thin long dashed line}. In \citet{hasinger08}, luminosity
dependence of the fraction is derived as a function of redshift.
We use the relation derived at $z=1.2-1.6$. Although the corrected
luminosity function still has lower number density than the
broad-line $+$ narrow-line AGN luminosity function, the large
fraction of the discrepancy between the broad-line luminosity
function and total luminosity function seems to be explained
with the obscured fraction down to $L_{\rm 2-10~keV}=10^{43}$ erg s$^{-1}$.
The remaining discrepancy can be caused by the possible incompleteness
of the broad-line AGN BHMF due to the spectroscopic incompleteness
and the difficulties in identifying broad-line AGNs with substantial
host contamination.
The uncertainty of the fraction of narrow-line AGN is still large
and the sample size of the SXDS is limited.
Larger sample is necessary to understand the remaining discrepancy.

\section{Discussion}\label{sec_DISCUSS}

\subsection{Broad-line AGN BHMF at z $\sim$1.4 and Its evolution to $z=0$}

\begin{figure*}
 \begin{center}
  \includegraphics[scale=1.15]{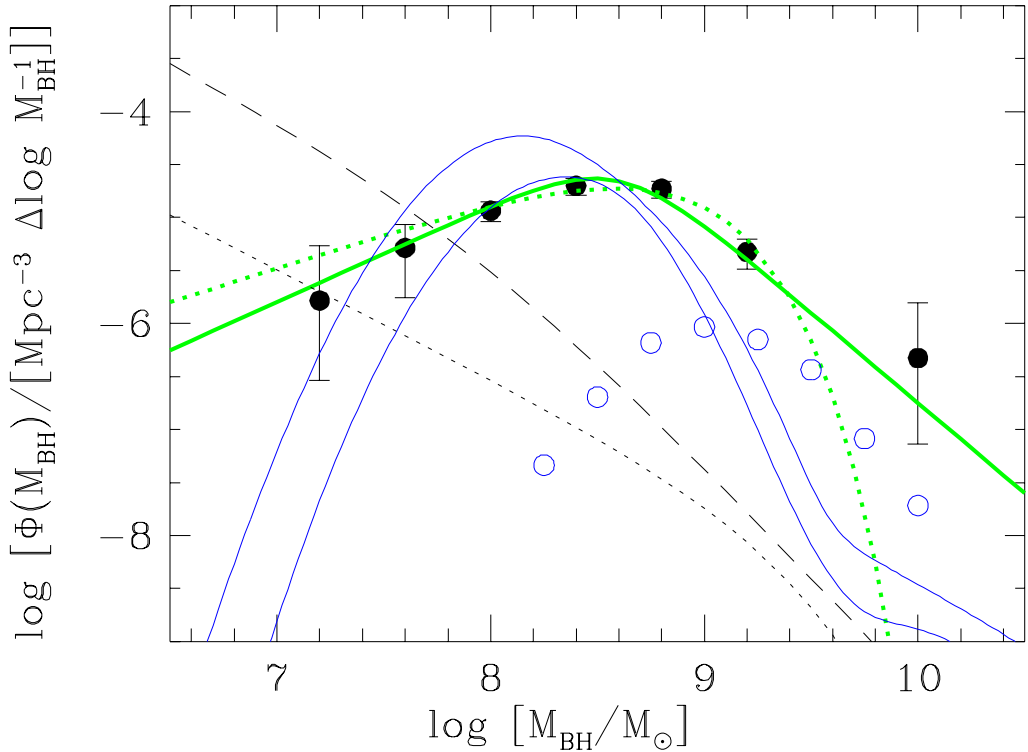}
  \caption{{\it Binned} ({\it filled circles}) and {\it corrected} 
broad-line AGN BHMFs
({\it thick solid line} with double-power-law model and {\it thick dashed line} with
Schechter model) from SXDS compared with {\it binned} ({\it open circles}) 
and {\it estimated} ({\it thin solid lines}) broad-line AGN
BHMFs at $z\sim1.4$ from SDSS \citep{shen12}.
Upper and lower {\it thin solid lines} enclose the 68\% possible
area.
The {\it thin dashed line} shows the {\it corrected} broad-line AGN BHMF 
in the local Universe from \citet{schulze10}.
The {\it thin dotted line} indicates the total BHMF at $z=6$ from
\citet{willott10}.
  \label{mass_plot2}}
 \end{center}
\end{figure*}

The {\it binned} and {\it corrected} broad-line AGN
BHMFs from the soft-band sample
are compared with the 
{\it binned} and {\it estimated} broad-line AGN BHMF from the
SDSS sample \citep{shen12} in Figure~\ref{mass_plot2}. The SXDS {\it binned}
BHMFs ({\it filled circles}) is larger than the {\it binned} BHMF
of the SDSS sample ({\it open circles}) by 1 order of magnitude at 
$10^{8.5} M_{\sun}$, and by two orders of magnitude at $10^8 M_{\sun}$. 
This is mostly because broad-line AGNs with lower Eddington ratio are detected 
in the SXDS sample than in the SDSS; the SDSS sample is only 30\% complete 
down to $M_{\rm BH}$ of $\sim 10^{9} M_{\sun}$ and $\lambda_{\rm Edd}$ 
of $\sim0.6$ at $z=1.4$ \citep{kelly10}. Additionally, as discussed 
in Section~\ref{sec_UVOpt}, the SXDS sample covers even mildly 
obscured broad-line AGNs that may be missed in the SDSS selection 
of broad-line AGNs due to color and stellarity issues. The thin solid lines 
in the figure represent the upper and lower envelopes of the {\it estimated}
broad-line AGN
BHMF with a Bayesian approach \citep{shen12}. The estimated number 
density at $10^{8} M_{\sun}$ is consistent with the {\it binned} and 
{\it corrected} BHMF of SXDS.

We examine the evolution of broad-line AGN
BHMF from $z=1.4$ to $0$ by comparing the $z\sim1.4$
{\it corrected} BHMF with that in the local Universe. The {\it dashed lines} in the figure
show the corrected local ( $z<0.3$ ) broad-line AGN BHMF from \citet{schulze10}.
The {\it binned} BHMFs of broad-line AGNs in the local Universe of \citet{greene09} 
and \citet{vestergaard09} are consistent with the {\it binned} BHMF of \citet{schulze10}. 
The {\it corrected} BHMF at $z\sim1.4$ exceeds that of local Universe in the mass
range above $10^{8} M_{\sun}$. However the behavior in the lower mass range is
different; the local broad-line AGN BHMF from \citet{schulze10} 
shows a steep increase with decreasing
black hole mass even below $10^{7} M_{\sun}$ mass range. On the contrary, 
the SXDS sample shows a hint of turn-over at a 
mass of $10^{8.5} M_{\sun}$. The difference in the
{\it corrected} broad-line AGN
BHMF may be indicative of a down-sizing trend of accretion activity
among the SMBH population. 
As mentioned in Section~\ref{sec_correction},
it needs to be noted that the identification
of broad-line AGNs could be incomplete in the low-mass end.

It is also possible that decline of the activity in the low-mass range 
from $z=0$ to $1.4$ is caused by luminosity and redshift dependent 
obscuration to the nucleus. Although broad-line AGNs have a wide range of $\lambda_{\rm Edd}$,
the fraction of broad-line AGNs with low-mass SMBH is higher in the lower luminosity
range. A luminosity dependence of the obscured fraction
is observed in various AGN samples \citep[e.g.][]{akiyama00, ueda03, simpson05, brusa10}; the obscured
fraction is higher among lower luminosity AGNs. Recently,
based on large samples of X-ray selected AGNs, it has also been suggested that
the fraction of obscured AGNs is higher at higher redshifts
\citep[e.g.][]{hasinger08, hiroi12}. Such redshift and luminosity dependent
obscuration may hide activity among low-mass SMBHs at high-redshifts.
We examine the contribution from obscured narrow-line AGNs to the
total active BHMF in Section~\ref{sec_ObscuredAGN}.

\subsection{Comparison with total BHMF at z $\sim$6}

In order to examine the growth of SMBHs at higher
redshift, we compare {\it binned} and {\it corrected} broad-line AGN
BHMF at $z=1.4$ 
with that at $z=6$. \citet{willott10} examined the total BHMF 
at $z\sim6$ using the optical luminosity function of 
broad-line AGNs at that redshift. They derive the total BHMF from the 
luminosity function using an Eddington ratio distribution 
derived from a part of their sample. They estimate the total 
(active + non-active) BHMF by crudely correcting the obscured 
fraction and duty cycle (active fraction). The resulting 
best-estimate total BHMF is shown with the {\it thin dotted line} 
in Figure~\ref{mass_plot2}. They argue the total BHMF is 
constrained down to $10^{8} M_{\sun}$. At that mass, 
the $z=6$ BHMF has a density of $2.5\times10^{-7}$ Mpc$^{-3}$. 
The number density matches the number density of $z=1.4$ corrected 
BHMF of broad-line AGNs at $10^{9.5} M_{\sun}$. If we naively assume
that the $10^{8} M_{\sun}$ black holes at $z=6$ grow to the $10^{9.5} M_{\sun}$
black holes at $z=1.4$ through accretion, then this implies mass growth by a factor 30
between $z=6$ and $1.4$, i.e., in 3.5 Gyr period. Such growth
implies that the multiple of the Eddington ratio and duty cycle has
order of 0.1 in that period for the massive SMBHs, if we use
the rest mass energy to radiation energy conversion efficiency of 0.1.
It needs to be noted that the high-mass end of the
{\it corrected} BHMF can be affected by the flattening due to the uncertainties
of the virial black hole mass estimates, therefore the effect needs to
be corrected before further quantitative evaluation of the growth of 
SMBHs between $z=6$ to $1.4$.

\subsection{Eddington Ratio Distribution Functions}

\begin{figure*}
 \begin{center}
  \includegraphics[scale=1.15]{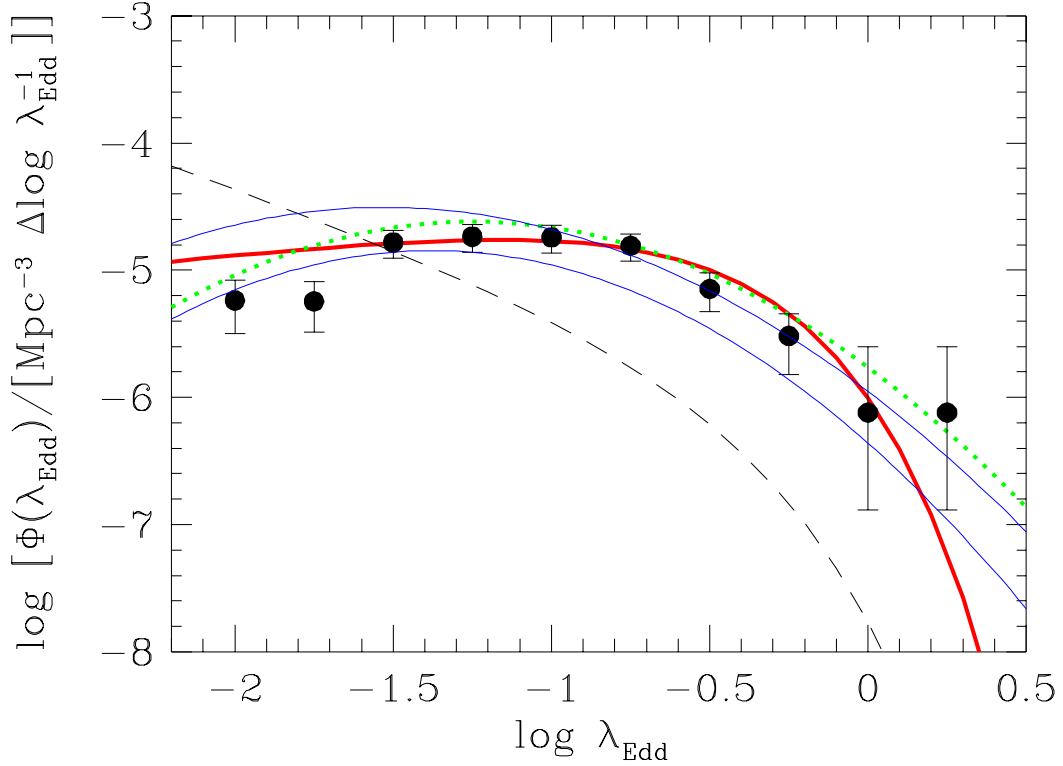}
  \caption{{\it Binned} ({\it filled circles}) and {\it corrected} broad-line AGN ERDF 
({\it thick solid line} with Schecheter ERDF and {\it thick dotted line} with log-normal ERDF) 
from SXDS compared with the
{\it estimated} broad-line AGN ERDF at $z\sim1.4$ from SDSS \citep{shen12}
({\it thin solid lines}). The upper and lower {\it thin solid lines}
represent the 68\% enclosed region.
The {\it thin dashed line} shows the {\it corrected} broad-line AGN
ERDF in the local
Universe from \citet{schulze10}.
  \label{ER_plot2}}
 \end{center}
\end{figure*}

Both of the {\it binned} and {\it corrected} broad-line AGN
ERDF show a decline at 
Eddington ratio of 1. Such a distribution suggests that
the accretion of X-ray-selected broad-line AGNs is limited by the Eddington
luminosity.
In Figure~\ref{ER_plot2}, we compare the {\it corrected} broad-line AGN
ERDF with
the {\it estimated} broad-line AGN ERDF at $z\sim1.4$ from the SDSS sample
({\it thin solid lines}, Shen \& Kelly 2012). 
The shapes of the ERDFs match rather well. 
It needs
to be noted that there is a possibility that we miss broad-line
AGNs accreting at a rate close to the Eddington limit because our 
X-ray selection is in relatively high energy range; the energy 
range of the soft-band sample corresponds to 1.2--4.8~keV at $z=1.4$. 
Optical to X-ray SED models of broad-line AGNs predict weak X-ray emission 
with steep X-ray spectrum among broad-line AGNs accreting with close to the 
Eddington limit \citep{kawaguchi01, done12}.

In the same figure, we also plot the {\it corrected} broad-line AGN
ERDF in the local 
Universe from \citet{schulze10}. The shape of their {\it binned}
ERDF in the local universe is consistent with that derived
by \citet{kauffmann09} for a narrow-line AGN sample from SDSS.
The shape of the {\it corrected} ERDF in the local Universe has a knee at a similar 
Eddington ratio
($\log \lambda_{\rm Edd}$ of $-0.6$; Schulze \& Wisotzki 2010)
to that of the {\it corrected} ERDF at $z=1.4$,
but shows a steeper increase in the low Eddington ratio range
down to $\log \lambda_{\rm Edd}$ of $-2.0$. 
The evolution of the {\it corrected} broad-line AGN
ERDF from $z=1.4$ to $0$ indicates the 
fraction of broad-line AGNs with $\lambda_{\rm Edd}$ close to 1 is higher 
at higher redshifts.

\subsection{Contribution from Obscured Narrow-line AGNs}\label{sec_ObscuredAGN}

\begin{figure}
 \begin{center}
  \includegraphics[scale=0.85]{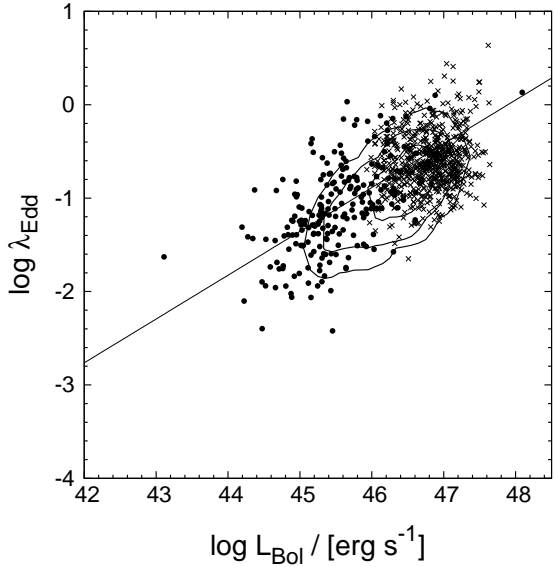}
  \caption{Bolometric luminosity versus Eddington ratio of SXDS broad-line AGNs
({\it filled circles}). The {\it solid line} represents
the least square fitting result for the SXDS AGNs.
The distribution of SDSS DR5 broad-line AGNs whose $M_{\rm BH}$ is estimated
from \ion{Mg}{2} \citep{shen08a} is shown by the {\it contours}.
{\it Crosses} represent broad-line AGNs from LBQS \citep{vestergaard09}.
\label{fig_L-E}}
 \end{center}
\end{figure}

\begin{figure}
 \begin{center}
  \includegraphics[scale=0.65]{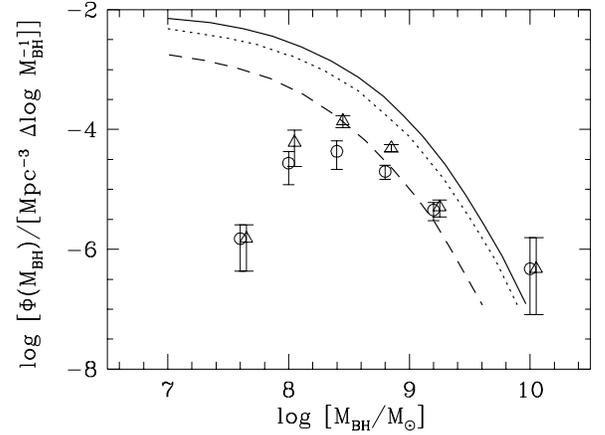}
  \caption{{\it Binned} active BHMF with ({\it open triangles}) and without ({\it open circles}) 
obscured narrow-line AGNs derived from the hard-band sample.
Solid, dotted, and dashed lines indicate the non-active BHMFs at $z=0$, $1$,
and $2$ derived with the $K$-band galaxy luminosity functions from \citet{li11}.
\label{fig_MFHC}}
  \end{center}
\end{figure}

\begin{deluxetable}{rrcrr}
\tabletypesize{\footnotesize}
\tablecaption{{\it Binned} active BHMF including Obscured Narrow-line AGNs \label{tbl_MF_AGN}}
\tablewidth{0pt}
\tablehead{
\multicolumn{1}{c}{$\langle \log [ M_{\rm BH} / M_{\sun}] \rangle$\tablenotemark{a}} &
\multicolumn{3}{c}{$\Phi_{\rm hard}( \log M_{\rm BH})$\tablenotemark{b}} &
\colhead{$N_{\rm hard}$} \\
}
\startdata
 7.2 & \multicolumn{3}{c}{\nodata} &  0 \\ 
 7.6 &   1.50 & $\pm$      &  1.07 &  2 \\
 8.0 &  60.95 & $\pm$      & 36.90 & 18 \\
 8.4 & 139.20 & $\pm$      & 31.42 & 87 \\
 8.8 &  48.81 & $\pm$      &  7.08 & 73 \\
 9.2 &   5.06 & $\pm$      &  1.60 & 10 \\
 9.6 & \multicolumn{3}{c}{\nodata} &  0 \\ 
10.0 &   0.47 & $^{+}_{-}$ & $^{1.09}_{0.40}$\tablenotemark{c} &  1 \\
\enddata
\tablenotetext{a}{The central value of $M_{\rm BH}$ in each bin. The bin size is 0.4 dex and extends  $\pm$ 0.2dex from the central value.}
\tablenotetext{b}{In unit of $10^{-6} \left[ {\rm{Mpc}}^{-3} \left( \Delta \log [ M_{\rm BH} / M_{\sun}] \right)^{-1} \right] $}
\tablenotetext{c}{The upper and lower limits are determined following Gehrels (1986).}
\end{deluxetable}

Among the X-ray-selected AGNs in the redshift range, more than half
of the AGNs are obscured narrow-line AGNs as shown in Table~\ref{tbl_MFsample}.
In this subsection, the contribution of these obscured narrow-line AGNs
to the {\it binned} active BHMF is evaluated using the hard-band sample which is less biased
against obscured AGNs than the soft-band sample. In the calculation of the 
{\it binned} active BHMF, not only spectroscopically-identified narrow-line AGNs
but also narrow-line AGNs with a photometric redshift are included.
Because we use the sample of
the hard X-ray selected AGNs,
the contribution from the heavily-obscured Compton-thick AGNs 
are not included. The black hole mass of the 
obscured narrow-line AGNs cannot be estimated from the FWHM of broad-line. 
Here, their black hole mass is estimated assuming they have same 
$\lambda_{\rm Edd}$ as broad-line AGNs with the same bolometric luminosity. 

First, the relation between the Eddington ratio and
the bolometric luminosity is determined for the broad-line AGNs in the 
SXDS sample.
The $L_{\rm bol}$ vs. $\lambda_{\rm Edd}$ distribution of the broad-line
AGNs is shown in Figure~\ref{fig_L-E}. There is a correlation between
the Eddington ratio and the bolometric luminosity.
The {\it solid line} in the figure represents the relation 
determined with the least square fitting,
\begin{eqnarray}
  \log \lambda_{\rm Edd} = 0.469 \times \log L_{\rm bol} - 22.46. 
\end{eqnarray}
The relation is mainly driven by the virial black hole
mass estimator used, and the scatter reflects the distribution of the
FWHM. The scatter of the $\lambda_{\rm Edd}$ from the best fit
relation is 0.4 dex, which is roughly consistent with the width
of the distribution of the FWHM.
In the figure, the
distribution of SDSS \citep{shen08a} and 
Large Bright Quasar Survey (LBQS) \citep{vestergaard09}
broad-line AGNs are also shown with the {\it contours} and {\it crosses}.
For the SDSS sample, only broad-line AGNs 
with $M_{\rm BH}$ from \ion{Mg}{2} measurements are shown. 
All of the broad-line AGNs follow the same trend of increasing
$\log \lambda_{\rm Edd}$ with increasing $L_{\rm Bol}$ as discussed in
\citet{croom11}.

For obscured narrow-line AGNs, their bolometric luminosities are 
estimated from the absorption corrected 2--10~keV luminosity with a bolometric
correction factor for the 2--10~keV luminosity \citep{marconi04}. 
We do not use UV or optical
luminosities to estimate bolometric luminosity, because they  
are severely suffered from dust extinction and contamination from host galaxies.
Using the above Eddington ratios for three bolometric luminosity bins, 
we estimate black hole mass of each narrow-line AGNs as
\begin{eqnarray}
  M_{\rm BH} [M_{\sun}] = \frac{L_{\rm bol}[{\rm erg \ s^{-1}}]}{1.26 \times 10^{38} \times \lambda_{\rm Edd, median}}. 
\end{eqnarray}
The scatter of the $L_{\rm bol}$ and $\lambda_{\rm Edd}$ relation is large,
thus the black hole mass estimation is only valid in statistical sense.
The {\it binned} BHMF derived including the narrow-line AGNs is shown in Figure~\ref{fig_MFHC}
and is tabulated in Table~\ref{tbl_MF_AGN}. The contribution of the obscured
narrow-line AGNs is large in the mass range, $\log M_{\rm BH} \leq 8.5$.

In order to examine the fraction of active black holes in the
entire SMBH population, we compare the {\it binned} active BHMF including narrow-line AGNs 
with non-active BHMF at intermediate redshifts derived from the galaxy luminosity 
functions and $M_{\rm BH}$ vs. $L_{\rm bulge}$ relationship \citep{tamura06, li11}.
\citet{li11} estimate the non-active BHMF up to $z\sim2$ from the galaxy $K$-band 
luminosity function, stellar mass function, and their redshift evolution. 
They assume an evolution of the $M_{\rm BH}$ vs. $L_{\rm bulge}$ relationship
as a function of redshift of the form $M_{\rm BH} / L_{\rm spheroid, K} \propto (1+z)^{1.4}$ 
following \citet{bennert10}. They also consider the redshift evolution of
the average bulge-to-total luminosity ratio, even though they do not include luminosity
dependence of the ratio. The resulting non-active BHMF is consistent with
that estimated in the local Universe \citep{vika09}.
In the figure, the estimated non-active BHMFs at $z=0$, $1$, and $2$ from \citet{li11}
are plotted with the {\it solid line}, {\it dashed line}, and {\it dotted line}, 
respectively. The {\it binned} active BHMF including narrow-line AGNs lies between the
non-active BHMFs at $z=1$ and $2$ in the mass range above $M_{\rm BH}=10^{8.5} M_{\sun}$;
a rather high fraction of active SMBHs in the mass range is implied at $z=1.4$.
It needs to be noted that the {\it binned} active BHMF is not
corrected for the scatter of the virial black hole mass estimate 
and that of the relation between the $\log L_{\rm bol}$ and the $\log \lambda_{\rm Edd}$.
Therefore, the number density can be overestimated in the high-mass end
due to the contamination from the objects in the lower mass range. In order to
discuss the fraction of active SMBHs quantitatively, the scatters needs to 
be corrected.

\section{Summary}\label{sec_SUMMARY}

Black hole masses of X-ray-selected broad-line AGNs detected 
in the SXDS are estimated from the width of the \ion{Mg}{2} broad-line and the
3000{\AA} monochromatic luminosity.
Because optical spectroscopic observations covering the \ion{Mg}{2} wavelength
range are not complete for the entire X-ray-selected broad-line
AGNs, the width of
H$\alpha$ broad-line measured with NIR spectroscopic survey is 
also used to provide a supplementary estimate if black hole masses for broad-line AGNs whose 
\ion{Mg}{2} broad-line width is not available.
The sample of broad-line AGNs is selected using X-ray detection as well as
detection of a broad emission line of either \ion{Mg}{2} in the optical or 
H$\alpha$ in the NIR. Some
of the broad-line AGNs have red rest-frame UV-colors, suggesting 
mild obscuration to their nucleus. For such broad-line AGNs, the hard X-ray
luminosity is used to estimate their intrinsic 3000{\AA} monochromatic 
luminosity assuming typical optical to X-ray luminosity ratio
as a function of bolometric luminosity. In total, black hole masses
are estimated for 215 broad-line AGNs at redshifts between 0.5 and 2.3.
In the redshift range between 1.18 and 1.68, the black hole mass
estimate is highly complete thanks to the supplemental measurements
of the H$\alpha$ line width from the NIR spectra.

Using the black hole masses of broad-line AGNs in the redshift
range, {\it binned} broad-line AGN BHMF and ERDF are initially
calculated using the $V_{\rm max}$ method. The {\it binned} BHMF shows a 
peak at $10^{8.5} M_{\sun}$. The {\it binned} ERDF has a steep decline at $\lambda_{\rm Edd}$
of 1 and a rather flat distribution below the Eddington limit.
Because the sample is X-ray-flux limited, both the {\it binned} BHMF and
ERDF are affected by the detection limit; the low $M_{\rm BH}$ end of 
the {\it binned} BHMF misses the
low $\lambda_{\rm Edd}$ objects and the low $\lambda_{\rm Edd}$ end of the {\it binned} ERDF
does not include the low $M_{\rm BH}$ broad-line AGNs. The effect of the flux
limit is corrected by assuming that the ERDF is constant regardless of 
the black hole mass.
Applying the Maximum Likelihood method with
appropriate functional shapes for the broad-line AGN
BHMF and ERDF, we determine 
the {\it corrected} BHMF and ERDF. 
We do not correct for the effect of the 
uncertainties in the virial black hole mass estimates.
Thanks to the faint detection limit, the sample extends
down to $\lambda_{\rm Edd}$ of 0.01 in the redshift range, the
correction is rather small and the shapes of the {\it corrected}
BHMF and ERDF do not significantly differ from those of {\it binned}
BHMF and ERDF. 

The {\it corrected} broad-line AGN
BHMF peaks at $10^{8.5} M_{\sun}$ and shows
a possible decline at the low-mass end. The
number density around $10^{8} M_{\sun}$ is consistent
with that estimated from the SDSS sample. The shape of the
{\it corrected} BHMF is completely different to that in the
local Universe which shows a steep increase down to
$10^{7} M_{\sun}$. The evolution of the shape of the
{\it corrected}
BHMF of broad-line AGNs from $z=1.4$ to $z=0$ may be
indicative of a down-sizing trend of accretion 
activity among the SMBH population.

The {\it corrected} broad-line AGN ERDF also shows a strong decline at 
the Eddington-limit and rather flat distribution
below the limit. The shape is also consistent
with that estimated from the SDSS sample. The strong 
decline suggests that the accretion of X-ray-selected broad-line AGNs
is limited by the Eddington luminosity.
The shape of the {\it corrected} ERDF in the local Universe
has a knee at a similar Eddington ratio but shows a steeper
increase in the low Eddington ratio range down to $\lambda_{\rm Edd}$ of 0.01. 
The evolution of the ERDF from $z=1.4$ to $z=0$ indicates the
fraction of broad-line AGNs with $\lambda_{\rm Edd}$ close to 1 is higher
at higher redshifts.

Using the hard-band sample of X-ray-selected AGNs, 
we evaluate the contribution of obscured narrow-line AGNs to
the {\it binned} active BHMF. We estimate the black hole masses of narrow-line AGNs
assuming the same Eddington ratio as a function of 
bolometric luminosity as for broad-line AGNs.
Once the contribution
from narrow-line AGNs is included, the {\it binned} active BHMF shows
a comparable number density to the non-active BHMFs at $z=1-2$
above $M_{\rm BH}$ of $10^{8.5} M_{\sun}$. 
Such a high density of active SMBHs at $z=1.4$ suggests
that massive SMBHs have high active fraction in the redshift range.

\acknowledgments

We would like to thank the anonymous referee for the through
reviewing of the paper that helped improve the paper.
We would like to thank
Dr. Marianne Vestergaard for providing us the machine readable
version of the \ion{Fe}{2} template.
We warmly thank the staff members of the Subaru telescope for
their support during the observations. 
K.N and M.A. thank Dr. Toshihiro Kawaguchi for useful discussions.
This work is supported by JSPS Grant-in-Aid for Young Scientist (B) (18740118)
and Grant-in-Aid for Scientific Research (B) (21340042).

Facilities: \facility{Subaru(Suprime-Cam, FOCAS, FMOS)}, \facility{XMM-Newton}, \facility{UKIRT}, 
\facility{GALEX}, \facility{CFHT}, \facility{INT}, \facility{Keck}, \facility{VLT}, \facility{AAT},
\facility{Spitzer}, \facility{Magellan}.

\end{document}